\documentclass[prd,twocolumn,showpacs,superscriptaddress,nofootinbib,floatfix,showkeys,10pt]{revtex4-1}

\usepackage{graphicx}
\usepackage{amsmath}
\usepackage{bm}
\usepackage{yhmath}
\usepackage{mathtools}
\usepackage{wasysym}
\usepackage[colorlinks,citecolor=blue,urlcolor=blue,linkcolor=blue]{hyperref}
\usepackage{subfigure}
\usepackage{color}
\usepackage{cases}
\usepackage{subfigure}
\usepackage{times}
\usepackage{dcolumn,booktabs,bm}
\usepackage{slashed}
\usepackage{amsfonts,amssymb,stmaryrd,latexsym}
\usepackage{mathrsfs}
\usepackage{textcomp}
\usepackage{multirow}
\usepackage{cancel}
\usepackage{array}
\usepackage{enumerate,enumitem}
\usepackage{orcidlink}
\usepackage{dashrule}
\usepackage{multirow}
\usepackage{times}

\allowdisplaybreaks[4]
\newcounter{cnt}
\makeatletter
\let\oldhypertarget\hypertarget
\renewcommand{\hypertarget}[2]{%
  \oldhypertarget{#1}{#2}%
    \protected@write\@mainaux{}{%
        \string\expandafter\string\gdef
          \string\csname\string\detokenize{#1}\string\endcsname{#2}%
    }%
  }
\newcommand{\myhyperlink}[1]{%
  \hyperlink{#1}{\csname #1\endcsname}%
  }
\makeatother

\def\N2LO{{N$^2$LO}}

\newcommand{\rub}{\affiliation{Institut f\"ur Theoretische Physik II, Ruhr-Universit\"at Bochum, D-44780 Bochum, Germany }}
\newcommand{\pkusp}{\affiliation{School of Physics, Peking University, Beijing 100871, China}}
\newcommand{\pku}{\affiliation{School of Physics and Center of High Energy Physics, Peking University, Beijing 100871, China}}
\newcommand{\hbu}{\affiliation{College of Physics Science \& Technology, Hebei University, Baoding 071002, China}
\affiliation{Hebei Key Laboratory of High-precision Computation and Application of Quantum Field Theory, Baoding, 071002, China}
\affiliation{Hebei Research Center of the Basic Discipline for Computational Physics, Baoding, 071002, China}}

\begin{document}


\title{Double pole structures of $X_1(2900)$ as the $P$-wave $\bar{D}^*K^*$ resonances}

\author{Jun-Zhang Wang\,\orcidlink{0000-0002-3404-8569}}\email{wangjzh2022@pku.edu.cn}\pku

\author{Zi-Yang Lin\,\orcidlink{0000-0001-7887-9391}}\email{lzy\_15@pku.edu.cn}\pkusp

\author{Bo Wang\,\orcidlink{0000-0003-0985-2958}}\email{wangbo@hbu.edu.cn}
\hbu

\author{Lu Meng\, \orcidlink{0000-0001-9791-7138}}\email{lu.meng@rub.de}\rub
\author{Shi-Lin Zhu\,\orcidlink{0000-0002-4055-6906}}\email{zhusl@pku.edu.cn}\pku

\begin{abstract}
We reveal the double pole structures of the manifestly exotic tetraquark state $X_1(2900)$ in the scenario of $P$-wave $\bar{D}^*K^*$ dimeson resonance. We find that the observed enhancement signal associated with $X_1(2900)$ in $B^+ \to D^+D^-K^+$ by LHCb contains two $P$-wave poles denoted as $T_{cs1-}(2900)$ and $T^{\prime}_{cs1-}(2900)$, respectively. After considering the channel couplings among the $\bar{D}K$, $\bar{D}^*K$, $\bar{D}K^*$ and $\bar{D}^*K^*$ and the width of the $K^*$ meson, the masses and widths of the $S$-wave pole $T_{cs0+}(2900)$ and two $P$-wave poles $T_{cs1-}(2900)$ and $T^{\prime}_{cs1-}(2900)$ coincide with those of the $X_0(2900)$ and $X_1(2900)$ remarkably, which provides strong support for identifying $X_0(2900)$ and $X_1(2900)$ as  $\bar{D}^{(*)}K^{(*)}$ dimeson states. Furthermore, we extensively calculate all $S$-wave and $P$-wave $\bar{D}^{(*)}K^{(*)}$ systems up to $J=3$ and predict four new isoscalar charmed-strange dimeson-type tetraquark states: an $S$-wave state $T_{cs1+}(2900)$ with quantum number $J^P=1^+$, three $P$-wave states $T_{cs1-}(2760)$ with $J^P=1^-$, $T_{cs0-}(2760)$ with $J^P=0^-$ and $T_{cs2-}(2900)$ with $J^P=2^-$. These near-threshold poles can be searched for at LHCb, Belle II and BESIII. 
\end{abstract}

\maketitle

\section{Introduction}\label{sec:intro}



In the past two decades, the experimental discovery of an increasing number of near-threshold hadron states in heavy-flavor realm indicates the existence of exotic heavy-flavor hadronic molecules (see the reviews~\cite{Chen:2016qju,Guo:2017jvc,Liu:2019zoy,Lebed:2016hpi,Esposito:2016noz,Brambilla:2019esw,Chen:2021ftn,Chen:2022asf,Meng:2022ozq} for relevant progress). These new hadrons usually cannot be accommodated in the conventional meson and baryon spectroscopy predicted by the quark model due to their common near-threshold features, the abnormal decay behaviors or the exotic quantum numbers. Therefore, they provide a unique platform to study the non-perturbative behavior of the strong interaction.

Among the potential exotic mesons, the most representative near-threshold states include the charmoniumlike $X(3872)$, $Z_c(3900)$ and doubly charmed tetraquark $T_{cc}^+(3875)$, which were first observed in the hidden-charm channel $J/\psi \pi^+\pi^-$ of the $B$ meson decay~\cite{Belle:2003nnu}, the charged channel $J/\psi \pi^{\pm}$ of the $e^+e^- \to J/\psi \pi^+\pi^-$~\cite{BESIII:2013ris} and the invariant mass spectrum of $D^0D^0\pi^+$ from the prompt production of the proton-proton collision~\cite{LHCb:2021vvq,LHCb:2021auc}, respectively. An intriguing feature of these three states is that their masses are very close to the threshold of $D\bar{D}^{\ast}$/$DD^{\ast}$, especially for the $X(3872)$, whose mass difference relative to the neutral channel obtained by a Breit-Wigner fit to the line shape~\cite{Workman:2022ynf} can even reach
\begin{eqnarray}
    \delta_{BW}=m_{D^0}+m_{D^{0*}}-m_{X(3872)}=0.00\pm 0.18~\textrm{MeV}.  
\end{eqnarray}
This phenomenon implies that the  $X(3872)$, $Z_c(3900)$ and $T_{cc}^+(3875)$ do not emerge by accident. They can be correlated with each other by the interaction between the charmed meson pairs. In addition, there have been significant progress in the search of the exotic baryons. A typical example is the observation of three hidden-charm pentaquark states, the $P_c(4312)$, $P_c(4440)$, and $P_c(4457)$ in the $\Lambda_b \to J/\psi p K$ by the LHCb Collaboration~\cite{LHCb:2019kea}. The first and last two $P_c$ states 
lie about several to tens MeVs below the $\Sigma_c\bar{D}$ and $\Sigma_c\bar{D}^\ast$ thresholds, respectively~\cite{Workman:2022ynf}. These exotic hadrons are necessarily related to the corresponding hadron-hadron dynamics, so are excellent candidates for the hadronic molecules ~\cite{Chen:2016qju,Guo:2017jvc,Liu:2019zoy,Lebed:2016hpi,Esposito:2016noz,Brambilla:2019esw,Chen:2021ftn,Chen:2022asf,Meng:2022ozq}.

In 2020, the LHCb Collaboration observed two charmed-strange tetraquark states $X_{0}(2900)$ and $X_{1}(2900)$  (also named as $T_{cs0}(2900)$ and $T_{cs1}(2900)$ in the new naming scheme for the exotic hadrons~\cite{Gershon:2022xnn}) with the manifestly exotic quark content $\bar{c}\bar{s}ud$ in the $D^-K^+$ invariant mass spectrum of the $B^+\to D^+D^-K^+$~\cite{LHCb:2020bls,LHCb:2020pxc}. Their resonance parameters are summarized as
\begin{eqnarray}
X_0(2900): m=2866\pm7~\mathrm{MeV}, \Gamma=57\pm3~\mathrm{MeV}; \\
X_1(2900): m=2904\pm5~\mathrm{MeV}, \Gamma=110\pm12~\mathrm{MeV}.
\end{eqnarray}
This is the first observation of the manifestly exotic hadrons with open heavy flavor and has inspired great interest to reveal their nature~\cite{Chen:2020aos,He:2020btl,Liu:2020nil,Hu:2020mxp,Agaev:2020nrc,Wang:2021lwy,Ortega:2023azl,Wang:2023hpp,Chen:2023syh,Karliner:2020vsi,He:2020jna,Wang:2020xyc,Zhang:2020oze,Wang:2020prk,Lu:2020qmp,Tan:2020cpu,Albuquerque:2020ugi,Agaev:2022eeh,Liu:2020orv,Burns:2020epm}. Very recently, the LHCb Collaboration confirmed the existence of $X_{0}(2900)$ and $X_{1}(2900)$ and their $J^{P}$ quantum numbers in a different production channel $B^+\to D^{*\pm}D^{\mp}K^+$ \cite{LHCb:2024vfz}. The favored $J^P$ quantum numbers of the $X_{0}(2900)$ and $X_{1}(2900)$ are $0^+$ and $1^-$~\cite{LHCb:2020pxc,LHCb:2024vfz}, respectively. So it is natural to interpret $X_{0}(2900)$ as an $S$-wave molecular bound state of $\bar{D}^\ast K^\ast$ since its mass is close to the threshold of $\bar{D}^\ast K^\ast$~\cite{Molina:2010tx,Chen:2020aos,He:2020btl,Liu:2020nil,Hu:2020mxp,Agaev:2020nrc,Wang:2021lwy,Ortega:2023azl,Wang:2023hpp,Chen:2023syh}. Other explanations of $X_{0}(2900)$ include the $S$-wave compact $\bar{c}\bar{s}ud$ tetraquark~\cite{Karliner:2020vsi,He:2020jna,Wang:2020xyc,Zhang:2020oze,Wang:2020prk,Lu:2020qmp,Tan:2020cpu,Albuquerque:2020ugi,Agaev:2022eeh} and the kinematic effects from the triangle singularities~\cite{Liu:2020orv,Burns:2020epm}.  It is worth mentioning that the calculations based on the quark model~\cite{Wang:2020prk,Lu:2020qmp} does not support the explanation of isoscalar compact tetraquark $\bar{c}\bar{s}ud$, whose mass is lower than the experimental mass of the $X_0(2900)$. 

For the $1^-$ state, it seems hard to understand the approximate mass degeneracy between the $X_1(2900)$ with odd-parity and $X_{0}(2900)$ with even-parity. Consequently, the theoretical discussion on the inner structure of $X_1(2900)$ is still relatively lacking at present. Several possible interpretations include the $S$-wave molecular bound state of $\bar{D}_1(2420) K$ \cite{He:2020btl}, the $P$-wave excitation of the $\bar{D}^\ast K^\ast$ bound state \cite{Wang:2021lwy} and the orbitally excited compact tetraquark state \cite{He:2020jna}. 

A significant problem has not yet been answered, i.e., is it possible to depict the feature of $X_{0}(2900)$ and $X_{1}(2900)$ in a unified theoretical picture? In this work, we aim to answer this question and show that this ostensible mass degeneracy is a strong evidence for the dimeson nature of $X_{0}(2900)$ and $X_{1}(2900)$. In this scenario, the $X_{1}(2900)$ calls for a $P$-wave potential between $\bar{D}^\ast $ and $ K^\ast$, which has been thought to be difficult to generate hadronic molecules for a long time due to the repulsive contribution from the high partial wave centrifugal barrier \cite{He:2020btl,Liu:2020nil}. However, our recent work has clearly illustrated the mechanism of generating near-threshold $P$-wave resonance poles with a weakly attractive interaction \cite{Lin:2024qcq}. It is a universal result once the $P$-wave potential is weakly attractive rather than absolutely repulsive. An excellent candidate is the recently observed vector charmoniumlike state $Y(3872)$ in the $e^+e^- \to D\bar{D}$ by BESIII \cite{BESIII:2024ths}, which can be identified as the first $P$-wave $D\bar{D}^*$/$D^*\bar{D}$ resonance \cite{Lin:2024qcq}.

In this work, after reproducing the $X_{0}(2900)$ as an $S$-wave $\bar{D}^\ast K^\ast$ bound state, we find the existence of the near-threshold $P$-wave $\bar{D}^\ast K^\ast$ resonances with $J^P=1^-$ in the unified meson exchange model, which can be related to the $X_{1}(2900)$. More remarkably, the $X_{1}(2900)$ peak involve two underlying resonance poles denoted by $T_{cs1-}(2900)$ and $T_{cs1-}^{\prime}(2900)$, which are generated by the coupling among three different total spin channels of $\bar{D}^\ast K^\ast$ with $S=0,1,2$. Furthermore, the complete coupled-channel analysis involving the lower $\bar{D}K$, $\bar{D}^*K$ and $\bar{D}K^*$ channels indicates that the observed $X_{1}(2900)$ enhancement structure comes mainly from the $T_{cs1-}(2900)$ because of its stronger coupling to the $\bar{D}K$ channel.  We urge searching for the possible signal around 2.9 GeV dominated by the other resonance pole $T_{cs1-}^{\prime}(2900)$ in the final state of $\bar{D}K^*$ in $b$ decay processes such as $B^+ \to D^{(*)+}D^{(*)-}K^{*+}$ and the prompt production in proton-proton collision.

Furthermore, we perform extensive coupled-channel calculations for all $S$-wave and $P$-wave $\bar{D}^{(*)}K^{(*)}$ systems up to $J=3$, and predict more isoscalar charmed-strange dimeson-type tetraquarks. 

This paper is organized as follows. The theoretical framework for the $P$-wave $\bar{D}^*K^*$ resonances and the dimeson state explanation of $X_1(2900)$ are illustrated in Sec.~\ref{sec2}. The double pole structures of $X_1(2900)$ with the coupled-channel analysis are discussed in Sec.~\ref{sec3}. Predictions for the pole properties of more charmed-strange dimeson-type tetraquark states are presented in Sec.~\ref{sec4}. This paper ends with the summary and outlook in Sec.~\ref{sec5}.

\section{The dimeson state explanation of $X_1(2900)$: the P-wave $\bar{D}^*K^*$ resonance} \label{sec2}

\subsection{Lagrangians and effective potentials}

For the near-threshold dimeson hadrons composed of $\bar{D}^{*} K^{*}$, the typical scale--its center-of-mass momentum $\gamma_m$, is usually smaller than the strange quark mass $m_s^{\mathrm{QM}}\sim 500$ MeV and charm quark mass $m_c^{\mathrm{QM}}\sim 1600$ MeV in the constitute quark model~\cite{Silvestre-Brac:1996myf,Vijande:2004he}. In this case, the strange and charm quarks are approximately treated as spectators (heavy color source) in the scattering dynamics. So we can adopt heavy meson effective Lagrangians to describe the $\bar{D}^{*} K^{*}$ interaction. 
For the $\bar{D}^{*} K^{*}\to \bar{D}^{*} K^{*}$ scattering, there are $t$-channel interactions from exchanging the pseudoscalar mesons $\pi$, $\eta$,  vector mesons $\omega$, $\rho$ and scalar meson $\sigma$ in the meson exchange model. The relevant effective Lagrangians are constructed according to the heavy quark symmetry, chiral symmetry and the SU(2) flavor symmetry, i.e., \cite{Wise:1992hn,Yan:1992gz,Grinstein:1992qt,Casalbuoni:1996pg,Li:2012cs,Li:2012ss} 
\begin{eqnarray}
\mathcal{L}_{{\widetilde{\mathcal{D}}}^{(*)}{\widetilde{\mathcal{D}}}^{(*)}\sigma} &=& -2g_s{\widetilde{\mathcal{D}}}_b^{\dag}{\widetilde{\mathcal{D}}}_b\sigma+2g_s{\widetilde{\mathcal{D}}}_b^{*}\cdot{\widetilde{\mathcal{D}}}_b^{*\dag}\sigma,\label{lag1}\\
\mathcal{L}_{{\widetilde{\mathcal{D}}}^{(*)}{\widetilde{\mathcal{D}}}^{(*)}{P}} &=&+\frac{2g}{f_{\pi}}(\widetilde{\mathcal{D}}_b\widetilde{\mathcal{D}}_{a\lambda}^{*\dag}+\widetilde{\mathcal{D}}_{b\lambda}^{*}\widetilde{\mathcal{D}}_a^{\dag})\partial^{\lambda}P_{ba}+i\frac{2g}{f_{\pi}} \nonumber \\
&&\times v^{\alpha}\varepsilon_{\alpha\mu\nu\lambda}{\widetilde{\mathcal{D}}}_b^{*\mu}{\widetilde{\mathcal{D}}}_{a}^{*\lambda\dag}
\partial^{\nu}{P}_{ba},\\
\mathcal{L}_{{\widetilde{\mathcal{D}}}^{(*)}{\widetilde{\mathcal{D}}}^{(*)}{V}} &=& +\sqrt{2}\beta g_V{\widetilde{\mathcal{D}}}_b{\widetilde{\mathcal{D}}}_a^{\dag} v\cdot{V}_{ba}-2\sqrt{2}\lambda g_V \nonumber \\
&&\times \epsilon_{\lambda\mu\alpha\beta}v^{\lambda}(\widetilde{\mathcal{D}}_b\widetilde{\mathcal{D}}_a^{*\mu\dag}+\widetilde{\mathcal{D}}_b^{*\mu}\widetilde{\mathcal{D}}_a^{\dag})(\partial^{\alpha}V^{\beta}_{ba})\nonumber\\
   &&-\sqrt{2}\beta g_V{\widetilde{\mathcal{D}}}_b^*\cdot{\widetilde{\mathcal{D}}}_a^{*\dag}v\cdot{V}_{ba}\nonumber\\
   &&-i2\sqrt{2}\lambda g_V{\widetilde{\mathcal{D}}}_b^{*\mu}{\widetilde{\mathcal{D}}}_a^{*\nu\dag}
   \left(\partial_{\mu}{V}_{\nu}-\partial_{\nu}{V}_{\mu}\right)_{ba}, \\
\mathcal{L}_{{K}^{(*)}{K}^{(*)}\sigma} &=& -2g_s^{\prime}{K}_b^{\dag}{K}_b\sigma+2g_s^{\prime}{K}_b^{*}\cdot{K}_b^{*\dag}\sigma, \\
\mathcal{L}_{{K}^{(*)}{K}^{(*)}{P}} &=&+\frac{2g^{\prime}}{f_{\pi}}(K_bK_{a\lambda}^{*\dag}+K_{b\lambda}^{*}K_a^{\dag})\partial^{\lambda}P_{ba}+i\frac{2g^{\prime}}{f_{\pi}} \nonumber \\
&&\times v^{\prime\alpha}\varepsilon_{\alpha\mu\nu\lambda}{K}_b^{*\mu}{K}_{a}^{*\lambda\dag}
\partial^{\nu}{P}_{ba},\\
\mathcal{L}_{{K}^{(*)}{K}^{(*)}{V}} &=& +\sqrt{2}\beta^{\prime} g_V^{\prime}{K}_b{K}_a^{\dag} v^{\prime}\cdot{V}_{ba}-2\sqrt{2}\lambda^{\prime} g_V^{\prime} \nonumber \\
&&\times \epsilon_{\lambda\mu\alpha\beta}v^{\prime\lambda}(K_bK_a^{*\mu\dag}+K_b^{*\mu}K_a^{\dag})(\partial^{\alpha}V^{\beta}_{ba})\nonumber\\
   &&-\sqrt{2}\beta^{\prime} g_V^{\prime}{K}_b^*\cdot{K}_a^{*\dag}v^{\prime}\cdot{V}_{ba}\nonumber\\
   &&-i2\sqrt{2}\lambda^{\prime} g_V^{\prime}{K}_b^{*\mu}{K}_a^{*\nu\dag}
   \left(\partial_{\mu}{V}_{\nu}-\partial_{\nu}{V}_{\mu}\right)_{ba}, \label{lag2}
\end{eqnarray}
where the $g=0.59\pm 0.07\pm 0.01$ and $g^{\prime}=1.12\pm0.01$ are extracted from the decay widths of $D^*\to D\pi$ and $K^*\to K\pi$ \cite{Workman:2022ynf}, respectively, and the $f_{\pi}=132$ MeV is the pion decay constant. The $g_V=g_V^{\prime}=m_{\rho}/f_{\pi}=5.8$ \cite{Isola:2003fh,Bando:1987br}.  The $\beta=$0.9 is fixed by the vector meson dominance model, and the $\lambda=$ 0.56 GeV$^{-1}$ is determined through a comparison of the form factor between the theoretical calculation from the light cone sum rule and lattice QCD \cite{Isola:2003fh,Bando:1987br}. Here, the $\beta^{\prime}=0.835$ can be determined by the hidden-gauge symmetry of the vector meson \cite{Molina:2010tx}, which is very close to the value of $\beta$ in the charmed meson sector.  Therefore, we take  $\lambda^{\prime}=\lambda$ in this work and its impact on the numerical results is also discussed. For the coupling constant associated with the $\sigma$ meson,  $g_s=g_s^{\prime}=0.76$ \cite{Liu:2008xz}. The field matrices ${P}$ and ${V}$ in the SU(2) flavor symmetry  are written as
\begin{eqnarray}
{P} &=& \left(\begin{array}{cc}
\frac{\pi^0}{\sqrt{2}}+\frac{\eta}{\sqrt{6}} &\pi^+ \nonumber\\
\pi^- &-\frac{\pi^0}{\sqrt{2}}+\frac{\eta}{\sqrt{6}}  
\end{array}\right),\nonumber\\
{V} &=& \left(\begin{array}{cc}
\frac{\rho^0}{\sqrt{2}}+\frac{\omega}{\sqrt{2}} &\rho^+  \nonumber\\
\rho^- &-\frac{\rho^0}{\sqrt{2}}+\frac{\omega}{\sqrt{2}}  
\end{array}\right).\label{lag3}
\end{eqnarray}

Based on the above interaction vertices, the effective potentials for the $\bar{D}^{*}(p_1)K^{*}(p_2)\to \bar{D}^{*}(p_3)K^{*}(p_4)$ are summarized as follows
\begin{eqnarray}
&&\mathcal{V}^{\bar{D}^{*}K^{*}\to \bar{D}^{*}K^{*}}_{\sigma} =-g_sg_{s}^{\prime}\frac{(\overset{\rightarrow}{\epsilon_1}\cdot \overset{\rightarrow}{\epsilon_3}^{\dag})(\overset{\rightarrow}{\epsilon_2}\cdot \overset{\rightarrow}{\epsilon_4}^{\dag})}{\overset{\rightarrow}{q}^2+m_{\sigma}^2}C_{\sigma},  \\
&& \mathcal{V}^{\bar{D}^{*}K^{*}\to \bar{D}^{*}K^{*}}_{\pi/\eta}=\frac{gg^{\prime}}{f_{\pi}^2}\frac{((\overset{\rightarrow}{\epsilon_1}\times \overset{\rightarrow}{\epsilon_3}^{ \dag})\cdot\overset{\rightarrow}{q})((\overset{\rightarrow}{\epsilon_2}\times \overset{\rightarrow}{\epsilon_4}^{ \dag})\cdot\overset{\rightarrow}{q})}{\overset{\rightarrow}{q}^2+m_{\pi/\eta}^2} \nonumber \\
&& \quad \quad \quad \quad \quad \quad \quad \times 
C_{\pi/\eta},  \\
 && \mathcal{V}^{\bar{D}^{*}K^{*}\to \bar{D}^{*}K^{*}}_{\rho/\omega}
  =(\frac{\beta\beta^{\prime}g_Vg_V^{\prime}}{2}\frac{(\overset{\rightarrow}{\epsilon_1}\cdot \overset{\rightarrow}{\epsilon_3}^{\dag})(\overset{\rightarrow}{\epsilon_2}\cdot \overset{\rightarrow}{\epsilon_4}^{\dag})}{\overset{\rightarrow}{q}^2+m_{\rho/\omega}^2} \nonumber \\ && \quad \quad
+\frac{2\lambda\lambda^{\prime}g_Vg_V^{\prime}}{\overset{\rightarrow}{q}^2+m_{\rho/\omega}^2} [ (\overset{\rightarrow}{\epsilon_1}\cdot \overset{\rightarrow}{\epsilon_2})(\overset{\rightarrow}{\epsilon_3}^{\dag}\cdot \overset{\rightarrow}{q})(\overset{\rightarrow}{\epsilon_4}^{\dag}\cdot \overset{\rightarrow}{q})-(\overset{\rightarrow}{\epsilon_1}\cdot \overset{\rightarrow}{\epsilon_4}^{\dag}) \nonumber \\
  &&\quad \quad  \times(\overset{\rightarrow}{\epsilon_3}^{\dag}\cdot \overset{\rightarrow}{q})(\overset{\rightarrow}{\epsilon_2}\cdot \overset{\rightarrow}{q})-(\overset{\rightarrow}{\epsilon_2}\cdot \overset{\rightarrow}{\epsilon_3}^{\dag})(\overset{\rightarrow}{\epsilon_4}^{\dag}\cdot \overset{\rightarrow}{q})(\overset{\rightarrow}{\epsilon_1}\cdot \overset{\rightarrow}{q}) \nonumber \\
  &&\quad \quad+(\overset{\rightarrow}{\epsilon_3}^{\dag} \cdot \overset{\rightarrow}{\epsilon_4}^{\dag})(\overset{\rightarrow}{\epsilon_1}\cdot \overset{\rightarrow}{q})(\overset{\rightarrow}{\epsilon_2}\cdot \overset{\rightarrow}{q}) ])C_{\rho/\omega},   
  \end{eqnarray}
where the transferred momentum $\overset{\rightarrow}{q}=\overset{\rightarrow}{p_1}-\overset{\rightarrow}{p_3}$, and initial state momentum $|\overset{\rightarrow}{p_1}|=p$ and final state momentum $|\overset{\rightarrow}{p_3}|=p^{\prime}$. In our calculations, the iospin breaking effect is ignored and the isospin averaged masses of the charmed meson from the Particle Data Group (PDG) are taken \cite{Workman:2022ynf}.
The flavor wave function for the isoscalar $\bar{D}^{*}K^{*}$ system can be expressed as
\begin{eqnarray}
|0,0\rangle &=& \frac{1}{\sqrt{2}}(|K^{*0}\bar{D}^{*0}\rangle-|K^{*+}D^{*-}\rangle).
\end{eqnarray}
Hence, the corresponding isospin factors are $C_{\sigma}=1$, $C_{\pi}=-3/2$, $C_{\eta}=1/6$, $C_{\rho}=-3/2$ and $C_{\omega}=1/2$. The effective potentials of $\bar{D}^{*}(p_1)K^{*}(p_2)\to \bar{D}^{*}(p_3)K^{*}(p_4)$ are actually governed by the following six operators, i.e., 
\begin{eqnarray}
    \mathcal{O}_1&=&(\overset{\rightarrow}{\epsilon_1}\cdot \overset{\rightarrow}{\epsilon_3}^{\dag})(\overset{\rightarrow}{\epsilon_2}\cdot \overset{\rightarrow}{\epsilon_4}^{\dag}), \nonumber \\
    \mathcal{O}_2&=& ((\overset{\rightarrow}{\epsilon_1}\times \overset{\rightarrow}{\epsilon_3}^{ \dag})\cdot\overset{\rightarrow}{q})((\overset{\rightarrow}{\epsilon_2}\times \overset{\rightarrow}{\epsilon_4}^{ \dag})\cdot\overset{\rightarrow}{q}), \nonumber\\
    \mathcal{O}_3&=& (\overset{\rightarrow}{\epsilon_1}\cdot \overset{\rightarrow}{\epsilon_2})(\overset{\rightarrow}{\epsilon_3}^{\dag}\cdot \overset{\rightarrow}{q})(\overset{\rightarrow}{\epsilon_4}^{\dag}\cdot \overset{\rightarrow}{q}), \nonumber\\
    \mathcal{O}_4&=& (\overset{\rightarrow}{\epsilon_1}\cdot \overset{\rightarrow}{\epsilon_4}^{\dag})(\overset{\rightarrow}{\epsilon_3}^{\dag}\cdot \overset{\rightarrow}{q})(\overset{\rightarrow}{\epsilon_2}\cdot \overset{\rightarrow}{q}), \nonumber\\
     \mathcal{O}_5&=& (\overset{\rightarrow}{\epsilon_2}\cdot \overset{\rightarrow}{\epsilon_3}^{\dag})(\overset{\rightarrow}{\epsilon_4}^{\dag}\cdot \overset{\rightarrow}{q})(\overset{\rightarrow}{\epsilon_1}\cdot \overset{\rightarrow}{q}), \nonumber\\
      \mathcal{O}_6&=& (\overset{\rightarrow}{\epsilon_3}^{\dag} \cdot \overset{\rightarrow}{\epsilon_4}^{\dag})(\overset{\rightarrow}{\epsilon_1}\cdot \overset{\rightarrow}{q})(\overset{\rightarrow}{\epsilon_2}\cdot \overset{\rightarrow}{q}), \label{eq:operator1}
\end{eqnarray}
in which $ \mathcal{O}_1$, $ \mathcal{O}_2$, $ \mathcal{O}_3\sim \mathcal{O}_6$ refer to the scalar, pseudoscalar and vector meson exchange respectively. For the potentials $V_i= \mathcal{O}_{i}(p,p^{\prime},z) D(p,p^{\prime},z)$ with $z=\overset{\rightarrow}{p}\cdot\overset{\rightarrow}{p}^{\prime}/pp^{\prime}$ and $D(p,p^{\prime},z)=1/(p^2+p^{\prime2}-2pp^{\prime}z+m^2)$, the concrete expressions of their partial-wave expansion potentials under $S$-wave and $P$-wave are summarized in the Appendix.

\subsection{Complex scaling method}

In general, the generation of the $P$-wave bound state is more difficult compared with that of the $S$-wave bound state due to the repulsion effect of the centrifugal barrier. Even so, as indicated in our previous work \cite{Lin:2024qcq}, by continuously decreasing the coupling strength in the potential to be less attractive, the possible $P$-wave bound poles on the physical sheet tend to migrate into the unphysical sheet and become the $P$-wave resonance poles. Thus, in order to effectively search for the possible $P$-wave dimeson-type resonance, the complex scaling method is employed. 

The complex scaling method (CSM) is a powerful approach to simultaneously obtain the resonance and bound state poles by solving the eigenenergy in the complex scaled Schrödinger equation \cite{Aguilar:1971ve,Balslev:1971vb,Lin:2022wmj,Lin:2023ihj}. This is achieved by performing a complex rotation on the coordinate $\bm{r}$ or momentum $\bm{p}$ given by 
\begin{eqnarray}
    U(\theta)\bm{r}=\tilde{\bm r}=\bm{r}e^{i\theta}, ~~~U(\theta)\bm{p}=\tilde{\bm p}=\bm{p}e^{-i\theta}.
\end{eqnarray}
Based on this transformation, the complex scaled Schrödinger equation in the momentum space can be written as
\begin{eqnarray}
    E\phi(\tilde{\bm p})=\frac{\bm{p}^2e^{-i2\theta}}{2\mu}\phi(\tilde{\bm p})+\int \frac{d^3\bm ke^{-i3\theta}}{(2\pi)^3} V(\tilde{\bm p},\tilde{\bm k})\phi(\tilde{\bm k}), 
\end{eqnarray}
where $\mu$ is the reduced mass and $V(\tilde{\bm p},\tilde{\bm k})$ is the effective potential. For the resonance pole $E_{R}=M_R-i\Gamma_{R}/2$, we require the rotation angle $2\theta>|\mathrm{Arg}(E_R)|$ to ensure the valid analytical continuation of the resonance wave function. As long as this condition satisfied, the resonance pole position remains independent of the choice of the rotation angle. Additionally, in order to regularize the ultraviolet divergence in the above integral equation, we introduce a non-local monopole regulator 
\begin{eqnarray}
\mathcal{F}(p^2,p^{\prime2})= \frac{\Lambda^2}{\Lambda^2+p^2}\frac{\Lambda^2}{\Lambda^2+p^{\prime2}}
\end{eqnarray}
to suppress the potential contribution at the large momentum, where the cutoff $\Lambda$ is the only unknown parameter but can be reliably estimated by reproducing the $S$-wave $X_0(2900)$ pole associated with the $^1S_0$ isosinglet $\bar{D}^*K^*$ system.

\subsection{Results and discussion}

\begin{figure*}[!ht]
\begin{centering}
    \scalebox{1.0}{\includegraphics[width=\linewidth]{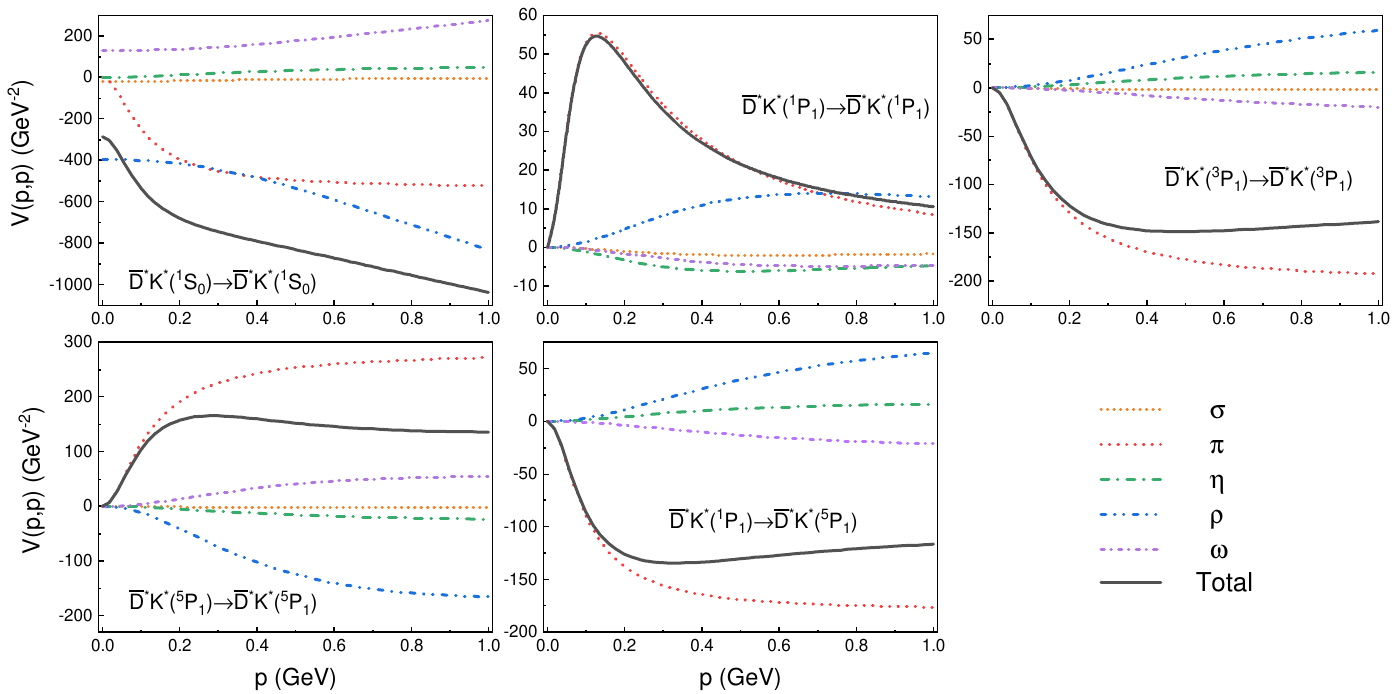}}    \caption{The partial-wave potentials of the isosinglet $\bar{D}^*K^*$ scattering  under $|^1S_0 \rangle \to |^1S_0 \rangle$, $|^1P_1 \rangle \to |^1P_1 \rangle$, $|^3P_1 \rangle \to |^3P_1 \rangle$, $|^5P_1 \rangle \to |^5P_1 \rangle$ and $|^1P_1 \rangle \to |^5P_1 \rangle$ in the meson exchange model. The $S$-wave channel and $P$-wave channels correspond to the $X_0(2900)$ and $X_{1}(2900)$, respectively.  Here, only $p=p^{\prime}$ results are shown. \label{fig:potential}}
\end{centering}
\end{figure*}

For the $\bar{D}^*K^*$ system with $J^{P}=1^-$, there are three partial-wave channels $|^{2S+1}L_{J}\rangle=|^1P_1\rangle$, $|^3P_1\rangle$ and $|^5P_1\rangle$, where $S$, $L$, $J$ denote the total spin, orbital angular momentum and total angular momentum, respectively. The $0^+$ $\bar{D}^*K^*$ system corresponds to the single channel $|^1S_0\rangle$. The partial-wave potentials of the isosinglet $\bar{D}^*K^*$ scattering in $|^1S_0 \rangle \to |^1S_0 \rangle$, $|^1P_1 \rangle \to |^1P_1 \rangle$, $|^3P_1 \rangle \to |^3P_1 \rangle$, $|^5P_1 \rangle \to |^5P_1 \rangle$ and $|^1P_1 \rangle \to |^5P_1 \rangle$ are depicted in Fig. \ref{fig:potential}. The potentials of the $^1S_0$ and $^3P_1$ channels are attractive and the potentials of the $^1P_1$ and $^5P_1$ channels are repulsive in total. For the nondiagonal interaction in the $1^{-}$ system, the coupling of $\bar{D}^*K^*(^1P_1) \to \bar{D}^*K^*(^5P_1) $ is very strong and other couplings between $|^1P_1/^5P_1\rangle$ and $|^3P_1\rangle$ are forbidden. There is a simple explanation of this selection rule. In the heavy meson limit, the $\bar{D}^*$ and $K^*$ act as the identical particles. The operators $\mathcal{O}_{1\sim3}$, $\mathcal{O}_{4}+\mathcal{O}_{5}$, $\mathcal{O}_{6}$ in Eqs. (\ref{eq:operator1}) are completely symmetric under the exchange operation of the $\bar{D}^*$ and $K^*$. However, the total spin wave functions of $|^1P_1\rangle$ and $|^5P_1\rangle$ have the same symmetry of particle exchange, but have the opposite symmetry with respect to the total spin wave function of $|^3P_1\rangle$. Thus, one concludes 
\begin{eqnarray}
\langle ^3P_1| \mathcal{O}_{1\sim3,6}/(\mathcal{O}_{4}+\mathcal{O}_{5})|^{1,5}P_1\rangle \equiv 0. \label{eq18}
\end{eqnarray}
Therefore, the three-channel calculations in the $1^-$ $\bar{D}^*K^*$  system is reduced to a two-channel plus single-channel scenario. From Fig. \ref{fig:potential}, it can also be found that the long-distance pion-exchange force dominates the $P$-wave interactions, whereas the $\rho$ exchange force is more important for the $S$-wave interaction. Therefore, it can be expected that the predictions for the possible $P$-wave poles should be robust due to the well-known pion exchange interaction.

\begin{figure}[t]
\begin{centering}
    \scalebox{1.0}{\includegraphics[width=\linewidth]{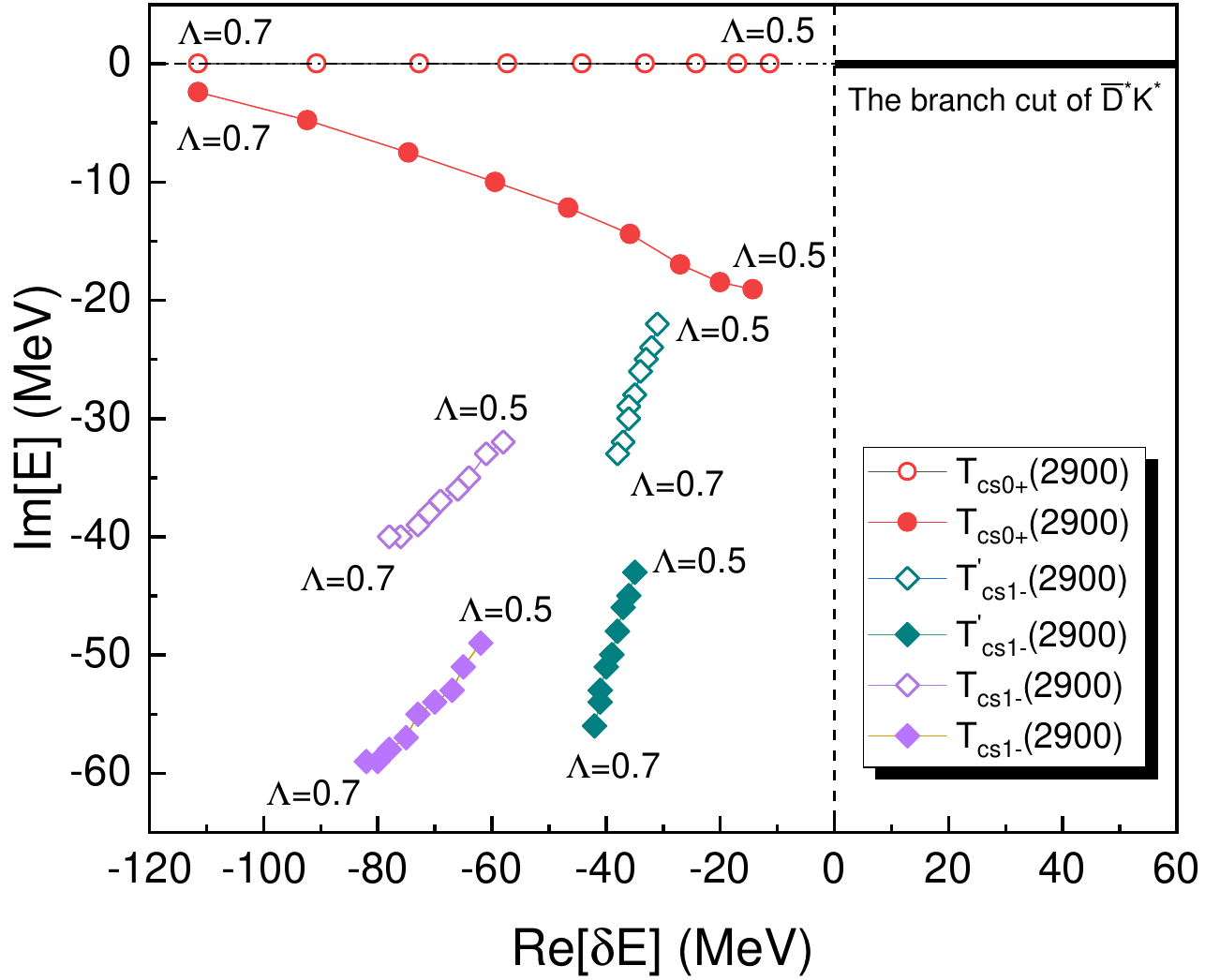}}
    \caption{ The pole trajectories of $S$-wave $T_{cs0+}(2900)$, $P$-wave  $T_{cs1-} (2900)$ and $T^{\prime}_{cs1-} (2900)$ with the varying cutoff parameter $\Lambda$ from 0.5 to 0.7 GeV. Here, the circle and diamond points correspond to the poles in the physical and unphysical Riemann sheets, respectively. The hollow and solid points represent the results without and with the width effect of the $K^*$ meson, respectively.  \label{fig:pole}}
\end{centering}
\end{figure}

For the $0^+$ $\bar{D}^*K^*$  system, we find a bound state pole denoted by $T_{cs0+}(2900)$, whose pole mass matches the experimental value of $X_0(2900)$ when the cutoff $\Lambda\approx0.57$ GeV. Intriguingly, within the same scattering dynamics, we find two near-threshold $P$-wave resonances with $J^P=1^-$ that are very close to each other in the unphysical Riemann sheet, in which the first resonance $T_{cs1-}(2900)$ is generated by the coupled-channel interaction between the $^1P_1$ and $^5P_1$ channels and the second resonance $T^{\prime}_{cs1-}(2900)$ corresponds to the $^3P_1$ channel.  In Fig. \ref{fig:pole}, we show the pole trajectories of $T_{cs0+}(2900)$, $T_{cs1-} (2900)$ and $T^{\prime}_{cs1-} (2900)$, as the cutoff $\Lambda$ varies from 0.5 to 0.7 GeV. One can see that the dependence of the $P$-wave states on the regulator is obviously insensitive compared to the $S$-wave case. Thus, the emergence of these two $P$-wave resonances is very natural so long as there exists an $S$-wave $0^+$ bound state.

The $X_0(2900)$ has a width of $57\pm3$ MeV \cite{LHCb:2020bls,LHCb:2020pxc}, which can be naturally understood in the $\bar{D}^*K^*$ molecule interpretation, if we consider the total width $\Gamma=51.4 \pm 0.8$ MeV~\cite{Workman:2022ynf} of the unstable $K^*$ meson. 
We introduce the energy-dependent width of $K^*$ ($D^*$ can be roughly considered as a stable particle here due to its tiny decay width) into the scattering dynamics of $\bar{D}^*K^* \to \bar{D}^*K^*$ by modifying the Schrödinger equation as \cite{Lin:2022wmj,Wang:2024ytk}
\begin{eqnarray}
    E\phi(\bm{p})=(\frac{\bm{p}^2}{2\mu}-i\frac{\Gamma(E)}{2})\phi(\bm{p})+\int \frac{d^3\bm{k}}{(2\pi)^3} V(\bm{p},\bm{{k}})\phi(\bm{k}), 
\end{eqnarray}
where $\Gamma(E)$ is the energy-dependent decay width of $K^* \to K \pi$. The pole positions of $T_{cs0+}(2900)$, $T_{cs1-} (2900)$ and $T^{\prime}_{cs1-} (2900)$ with the width of the $K^*$ are presented by the solid points in Fig. \ref{fig:pole}. The imaginary parts of the two $P$-wave poles are consistent with the large width of $X_1(2900)$ \cite{LHCb:2020bls,LHCb:2020pxc}, which strongly supports $X_1(2900)$ as the $P$-wave $\bar{D}^*K^*$ resonance.

\section{Coupled channel analysis for the double pole structures of $X_1(2900)$} \label{sec3}

In the above section, we reveal the novel double pole structures of $X_1(2900)$ in the $\bar{D}^*K^*$ resonance scenario. However, the enhancement structure associated with $X_1(2900)$ was observed in the invariant mass spectrum of $D^-K^+$. In order to clearly learn the contributions of two $P$-wave resonances in the $X_1(2900)$ structure, we need to carry out the complete coupled-channel analysis for $T_{cs1-} (2900)$ and $T^{\prime}_{cs1-} (2900)$, which involves six channels $\bar{D}K(^1P_1)$, $\bar{D}^*K(^3P_1)$, $\bar{D}K^*(^3P_1)$, $\bar{D}^*K^*(^1P_1)$, $\bar{D}^*K^*(^3P_1)$ and $\bar{D}^*K^*(^5P_1)$. Based on the effective Lagrangians in Eqs. (\ref{lag1})-(\ref{lag2}), the related effective potentials in the coupled-channel framework can be given as
\begin{eqnarray}
\mathcal{V}^{\bar{D}K^*\to \bar{D}K^*}_{\sigma} &=&-g_sg_{s}^{\prime}\frac{1}{\overset{\rightarrow}{q}^2+m_{\sigma}^2}C_{\sigma}, \\
\mathcal{V}^{\bar{D}K^*\to \bar{D}K^*}_{\rho/\omega} &=&\frac{\beta\beta^{\prime}g_Vg_V^{\prime}}{2}\frac{\overset{\rightarrow}{\epsilon_2}\cdot \overset{\rightarrow}{\epsilon_4}^{ \dag}}{\overset{\rightarrow}{q}^2+m_{\rho/\omega}^2}C_{\rho/\omega}, \\
\mathcal{V}^{\bar{D}^*K\to \bar{D}^*K}_{\sigma} &=&-g_sg_{s}^{\prime}\frac{1}{\overset{\rightarrow}{q}^2+m_{\sigma}^2}C_{\sigma}, \\
\mathcal{V}^{\bar{D}^*K\to \bar{D}^*K}_{\rho/\omega} &=&\frac{\beta\beta^{\prime}g_Vg_V^{\prime}}{2}\frac{\overset{\rightarrow}{\epsilon_1}\cdot \overset{\rightarrow}{\epsilon_3}^{ \dag}}{\overset{\rightarrow}{q}^2+m_{\rho/\omega}^2}C_{\rho/\omega}, \\
\mathcal{V}^{\bar{D}K\to \bar{D}K}_{\sigma} &=&-g_sg_{s}^{\prime}\frac{1}{\overset{\rightarrow}{q}^2+m_{\sigma}^2}C_{\sigma}, \\
  \mathcal{V}^{\bar{D}K\to \bar{D}K}_{\rho/\omega} &=&\frac{\beta\beta^{\prime}g_Vg_V^{\prime}}{2}\frac{1}{\overset{\rightarrow}{q}^2+m_{\rho/\omega}^2}C_{\rho/\omega}, \\
  \mathcal{V}^{\bar{D}^*K\to \bar{D}K^*}_{\pi/\eta} &=&\frac{-gg^{\prime}}{f_{\pi}^2}\frac{( \overset{\rightarrow}{\epsilon_1}\cdot\overset{\rightarrow}{q})( \overset{\rightarrow}{\epsilon_4}^{ \dag}\cdot\overset{\rightarrow}{q})}{\overset{\rightarrow}{q}^2+m_{\pi/\eta}^2}C_{\pi/\eta}, \\
   \mathcal{V}^{\bar{D}^*K\to \bar{D}K^*}_{\rho/\omega} &=&2\lambda\lambda^{\prime}g_Vg_V^{\prime}\frac{1}{\overset{\rightarrow}{q}^2+m_{\rho/\omega}^2}C_{\rho/\omega} \nonumber \\ 
   && \times [( \overset{\rightarrow}{\epsilon_1}\cdot\overset{\rightarrow}{q})( \overset{\rightarrow}{\epsilon_4}^{ \dag}\cdot\overset{\rightarrow}{q})-\overset{\rightarrow}{q}^2(\overset{\rightarrow}{\epsilon_1}\cdot\overset{\rightarrow}{\epsilon_4}^{\dag})],\\
  \mathcal{V}^{\bar{D}K\to \bar{D}^{*}K^{*}}_{\pi/\eta}
  &=& \frac{-gg^{\prime}}{f_{\pi}^2}\frac{( \overset{\rightarrow}{\epsilon_3}^{ \dag}\cdot\overset{\rightarrow}{q})( \overset{\rightarrow}{\epsilon_4}^{ \dag}\cdot\overset{\rightarrow}{q})}{\overset{\rightarrow}{q}^2+m_{\pi/\eta}^2}C_{\pi/\eta}, \label{eq28}\\
  \mathcal{V}^{\bar{D}K\to \bar{D}^{*}K^{*}}_{\rho/\omega}
  &=& 2\lambda\lambda^{\prime}g_Vg_V^{\prime}\frac{( \overset{\rightarrow}{\epsilon_3}^{ \dag}\times\overset{\rightarrow}{q})\cdot( \overset{\rightarrow}{\epsilon_4}^{ \dag}\times\overset{\rightarrow}{q})}{\overset{\rightarrow}{q}^2+m_{\rho/\omega}^2}C_{\rho/\omega}, \nonumber \label{eq29} \\ \\
   \mathcal{V}^{\bar{D}^*K\to \bar{D}^{*}K^{*}}_{\pi/\eta}
   &=& \frac{igg^{\prime}}{f_{\pi}^2}\frac{( \overset{\rightarrow}{\epsilon_4}^{ \dag}\cdot\overset{\rightarrow}{q})( \overset{\rightarrow}{\epsilon_3}^{ \dag}\times\overset{\rightarrow}{q})\cdot\overset{\rightarrow}{\epsilon_1}}{\overset{\rightarrow}{q}^2+m_{\pi/\eta}^2}C_{\pi/\eta}, \\
   \mathcal{V}^{\bar{D}^*K\to \bar{D}^{*}K^{*}}_{\rho/\omega}
  &=& 2i\lambda\lambda^{\prime}g_Vg_V^{\prime}\frac{( \overset{\rightarrow}{\epsilon_4}^{ \dag}\times\overset{\rightarrow}{q})\cdot\overset{\rightarrow}{\epsilon_3}^{ \dag}( \overset{\rightarrow}{\epsilon_1}\cdot\overset{\rightarrow}{q})}{\overset{\rightarrow}{q}^2+m_{\rho/\omega}^2}C_{\rho/\omega}, \nonumber \\ \\
   \mathcal{V}^{\bar{D}K^*\to \bar{D}^{*}K^{*}}_{\pi/\eta}
   &=& \frac{igg^{\prime}}{f_{\pi}^2}\frac{( \overset{\rightarrow}{\epsilon_3}^{ \dag}\cdot\overset{\rightarrow}{q})( \overset{\rightarrow}{\epsilon_4}^{ \dag}\times\overset{\rightarrow}{q})\cdot\overset{\rightarrow}{\epsilon_2}}{\overset{\rightarrow}{q}^2+m_{\pi/\eta}^2}C_{\pi/\eta}, \\
   \mathcal{V}^{\bar{D}K^*\to \bar{D}^{*}K^{*}}_{\rho/\omega}
  &=& 2i\lambda\lambda^{\prime}g_Vg_V^{\prime}\frac{( \overset{\rightarrow}{\epsilon_3}^{ \dag}\times\overset{\rightarrow}{q})\cdot\overset{\rightarrow}{\epsilon_4}^{ \dag}( \overset{\rightarrow}{\epsilon_2}\cdot\overset{\rightarrow}{q})}{\overset{\rightarrow}{q}^2+m_{\rho/\omega}^2}C_{\rho/\omega}. \nonumber \\
\end{eqnarray}

\begin{table}[htb]
\renewcommand{\arraystretch}{1.5}
\caption{The pole properties of the $S$-wave $T_{cs0+}(2900)$ in the coupled-channel calculations involving the $\bar{D}K(^1S_0)$ and $\bar{D}^*K^*(^1S_0)$ channels. \label{tab:x0}}
\setlength{\tabcolsep}{1.5mm}
{
\centering
\begin{tabular}{ccccccccc}
\toprule[0.8pt]
\toprule[0.8pt]
     $T_{cs0+}(2900)$ & $\lambda^{\prime}=0.56$  & $\lambda^{\prime}=0.28$ & $\lambda^{\prime}=0.84$  \\
     \hline
      Pole (GeV)   & 2.862-0.023i  & 2.878-0.029i &  2.840-0.014i \\
      \hline
      $\mathcal{P}(\bar{D}K)$    & (-0.5-1.8i) \% &(-0.6-1.9i) \% & (-0.3-1.5i) \%   \\
       \hline
       $\mathcal{P}(\bar{D}^*K^*)$    & (100.5+1.8i) \% &(100.6+1.9i) \% &  (100.3+1.5i) \%  \\
        \hline
         $Res(\bar{D}K)$    & 3.48 & 4.25 &  2.33  \\
       \hline
        $Res(\bar{D}^*K^*)$    & 136.0 & 85.7 &  205.3 \\
\bottomrule[0.8pt]
\bottomrule[0.8pt]
\end{tabular}
}
\end{table}

In the coupled-channel calculations, we fix the cutoff $\Lambda=0.6$ GeV and always consider the dynamical width effect of the unstable $K^*$ meson. We also investigate the impact of the unknown coupling constant $\lambda^{\prime}$ on the pole results by taking three typical values $\lambda^{\prime}=0.5\lambda, \lambda, 1.5 \lambda $. By solving the complex scaled coupled-channel Schrödinger equation, we can simultaneously obtain the poles and the corresponding wave functions in different channels.  For the non-Hermitian Hamiltonian, the wave function normalization should be defined by c-product \cite{Lin:2023ihj}, which reads 
\begin{eqnarray}
   ( \phi| \phi )=\sum_i\int\frac{d\bm{p}^3}{(2\pi)^3}e^{-3i\theta}\phi_i(\tilde{\bm p})^2=1,
\end{eqnarray}
where $\theta$ is the rotation angle in complex scaling method, and subscript $i$ stands for the involved channels. By employing the solved wave function $\phi(\bm p)$ in the momentum space, we can calculate the defined physical quantity $\mathcal{P}_i=( \phi_i |\phi_i )$, which roughly reflects the ratio of the $i$-th channel. Additionally, we can estimate the effective coupling strength between the dimeson pole and the $i$-th continuum channel by the residue of the scattering $T$-matrix or the pole wave function, specifically 
\begin{eqnarray}
Res(i)&=&g^2_i=\mathrm{lim}_{E\to E_R}(E-E_R)T_{ii}(E) \nonumber\\
    &=& |\langle k_{R,j} |\hat{V}|\phi\rangle|^2,
\end{eqnarray}
where $E_R$ is the pole position, $k_{R,j}=\sqrt{2\mu_i\delta E_{R,j}}$, and $T(E)$ is the on-shell $T$ matrix. Here, $\delta E_{R,j}=E_R-m_{\mathrm{th}}^i$ and $m_{\mathrm{th}}^i$ is the threshold of the $i$-th channel. One can find more details in Ref. \cite{Lin:2023ihj}.

\begin{figure}[t]
\begin{centering}
    \scalebox{1.0}{\includegraphics[width=\linewidth]{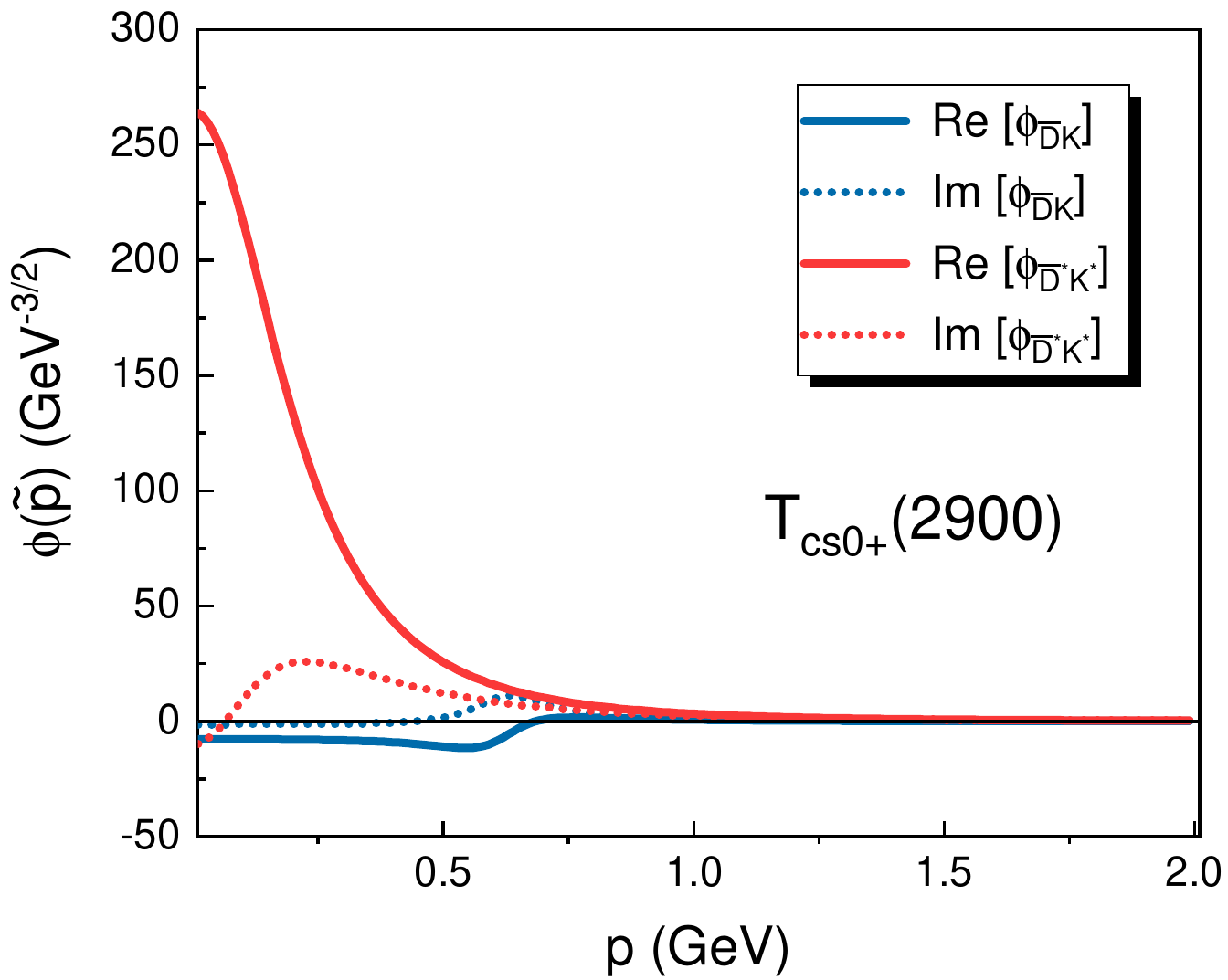}}
    \caption{The wave functions $\phi_{\bar{D}K}(\tilde{p})$ and $\phi_{\bar{D}^*K^*}(\tilde{p})$ of $S$-wave $T_{cs0+}(2900)$ with the rotation angle $\theta=10^{\mathrm{o}}$ in the coupled-channel calculations. The solid and dashed line stand for the real and imaginary part of the wave function, respectively.  \label{fig:wfs}}
\end{centering}
\end{figure}

\begin{table}[htb]
\renewcommand{\arraystretch}{1.5}
\caption{The pole properties of the $P$-wave $T_{cs1-}(2900)$ in the coupled-channel calculations involving the $\bar{D}K(^1P_1)$, $\bar{D}^*K(^3P_1)$, $\bar{D}K^*(^3P_1)$, $\bar{D}^*K^*(^1P_1)$, $\bar{D}^*K^*(^3P_1)$ and $\bar{D}^*K^*(^5P_1)$ channels. \label{tab:x1}}
\setlength{\tabcolsep}{1.2mm}
{
\centering
\begin{tabular}{ccccccccc}
\toprule[0.8pt]
\toprule[0.8pt]
     $T_{cs1-}(2900)$ & $\lambda^{\prime}=0.56$  & $\lambda^{\prime}=0.28$ & $\lambda^{\prime}=0.84$  \\
     \hline
      Pole (GeV)   & 2.840-0.049i  & 2.838-0.057i &  2.844-0.044i \\
      \hline
      $\mathcal{P}(\bar{D}K)$    & (0.2+0.7i) \% &(-0.0+0.8i) \% & (0.2+0.5i) \%   \\
       \hline
      $\mathcal{P}(\bar{D}^*K)$    & (0.5-0.2i) \% &(0.7-0.0i) \% & (0.5-0.2i) \%   \\
       \hline
       $\mathcal{P}(\bar{D}K^*)$    & (3.0-3.9i) \% &(5.3-3.6i) \% & (1.8-3.5i) \%   \\
       \hline
       $\mathcal{P}(\bar{D}^*K^*(^1P_1))$    & (67.5-2.6i) \% &(57.5-12.5i) \% &  (74.5-1.5i) \%  \\
        \hline
        $\mathcal{P}(\bar{D}^*K^*(^3P_1))$    & (7.2+11.3i) \% &(6.8+16.1i) \% &  (6.3+9.3i) \%  \\
       \hline
        $\mathcal{P}(\bar{D}^*K^*(^5P_1))$    & (21.6-5.3i) \% &(29.7-0.8i) \% &  (16.7-4.6i) \%  \\
       \hline
        $Res(\bar{D}^*K^*)(^1P_1)$    & 6.55 & 6.64 &  10.7  \\
       \hline
        $Res(\bar{D}^*K^*)(^3P_1)$    & 2.33 & 2.97 &  2.87  \\
       \hline
        $Res(\bar{D}^*K^*)(^5P_1)$    & 2.65 & 3.15 &  3.78  \\
       \bottomrule[0.8pt]
\bottomrule[0.8pt]
\end{tabular}
}
\end{table}

\begin{table}[tb]
\renewcommand{\arraystretch}{1.5}
\caption{The pole properties of the $P$-wave $T^{\prime}_{cs1-}(2900)$ in the coupled-channel calculations involving the $\bar{D}K(^1P_1)$, $\bar{D}^*K(^3P_1)$, $\bar{D}K^*(^3P_1)$, $\bar{D}^*K^*(^1P_1)$, $\bar{D}^*K^*(^3P_1)$ and $\bar{D}^*K^*(^5P_1)$ channels. \label{tab:x1prime}}
\setlength{\tabcolsep}{0.8mm}
{
\centering
\begin{tabular}{ccccccccc}
\toprule[0.8pt]
\toprule[0.8pt]
     $T^{\prime}_{cs1-}(2900)$ & $\lambda^{\prime}=0.56$  & $\lambda^{\prime}=0.28$ & $\lambda^{\prime}=0.84$  \\
     \hline
      Pole (GeV)   & 2.864-0.038i  &2.863-0.041i & 2.865-0.035i  \\
      \hline
      $\mathcal{P}(\bar{D}K)$    & (-0.0+0.0i) \% & (-0.1+0.0i) \%& (-0.0+0.0i) \%   \\
       \hline
      $\mathcal{P}(\bar{D}^*K)$    & (0.3+0.6i) \% & (0.3+0.7i) \%&  (0.3+0.5i) \%  \\
       \hline
       $\mathcal{P}(\bar{D}K^*)$    & (4.5+2.5i) \% &(4.7+3.6i) \% & (4.0+1.9i) \%   \\
       \hline
       $\mathcal{P}(\bar{D}^*K^*(^1P_1))$    & (1.7+4.5i) \% & (1.3+5.0i) \%& (1.8+4.2i) \%   \\
        \hline
        $\mathcal{P}(\bar{D}^*K^*(^3P_1))$    & (91.4-15.0i) \% &(93.1-18.3i) \% &  (91.5-12.7i) \%  \\
       \hline
        $\mathcal{P}(\bar{D}^*K^*(^5P_1))$    & (2.1+7.4i) \% &(0.7+9.0i) \% & (2.4+6.1i)\%   \\
       \hline
        $Res(\bar{D}^*K^*)(^1P_1)$    & 0.50 & 0.60 &  0.43  \\
       \hline
        $Res(\bar{D}^*K^*)(^3P_1)$    & 11.69 & 13.1 &  10.8  \\
       \hline
        $Res(\bar{D}^*K^*)(^5P_1)$    & 0.62 & 0.76 &  0.51  \\
      \bottomrule[0.8pt]
\bottomrule[0.8pt]
\end{tabular}
}
\end{table}

\begin{figure*}[ht]
\begin{centering}
    \scalebox{1.0}{\includegraphics[width=\linewidth]{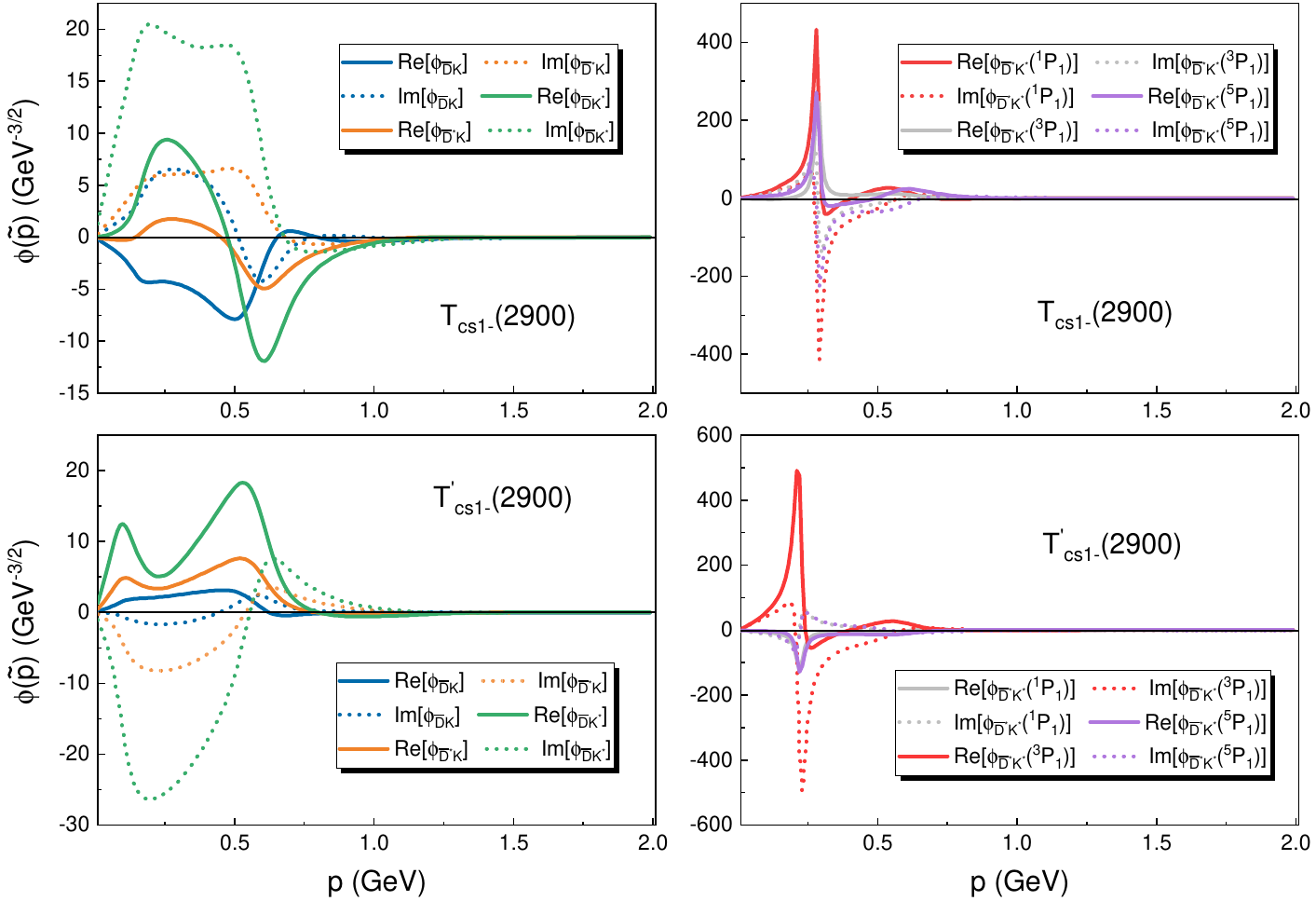}}
    \caption{The wave functions $\phi_{\bar{D}K}(\tilde{p})$, $\phi_{\bar{D}^*K}(\tilde{p})$, $\phi_{\bar{D}K^*}(\tilde{p})$, $\phi_{\bar{D}^*K^*(^1P_1)}(\tilde{p})$, $\phi_{\bar{D}^*K^*(^3P_1)}(\tilde{p})$ and $\phi_{\bar{D}^*K^*(^5P_1)}(\tilde{p})$  of two $P$-wave resonances $T_{cs1-}(2900)$ and $T^{\prime}_{cs1-}(2900)$  with the rotation angle $\theta=78^{\mathrm{o}}$ in the coupled-channel calculations. The solid and dashed lines stand for the real and imaginary part of the wave function, respectively. \label{fig:wfp}}
\end{centering}
\end{figure*}

The pole properties of $S$-wave $T_{cs0+}(2900)$ in the coupled-channel analysis are listed in Table \ref{tab:x0}. Accordingly, the wave functions $\phi_{\bar{D}K}(\tilde{p})$ and $\phi_{\bar{D}^*K^*}(\tilde{p})$ of $T_{cs0+}(2900)$ with the rotation angle $\theta=10^{\mathrm{o}}$ are depicted in Fig. \ref{fig:wfs}. When involving the lower $\bar{D}K$ channel, the quasi-bound state $T_{cs0+}(2900)$ can decay into the $\bar{D}K$ final state, so its pole width increases by 20 MeV. Both the mass and width of $T_{cs0+}(2900)$ coincide very well with the LHCb measurements for the $X_0(2900)$ \cite{LHCb:2020bls,LHCb:2020pxc} after considering the coupled-channel effect. In addition, because the vector meson exchanges dominate the interaction of $\bar{D}^*K^*(^1S_0) \to \bar{D}^*K^*(^1S_0) $, the properties of $T_{cs0+}(2900)$ are relatively sensitive to the coupling constant $\lambda^{\prime}$. Its binding becomes deeper and its decay width becomes smaller with increasing $\lambda^{\prime}$.

With a large rotation angle $\theta=78^{\mathrm{o}}$, two $P$-wave states $T_{cs1-}(2900)$ and $T^{\prime}_{cs1-}(2900)$ appear and the corresponding wave functions  $\phi_{\bar{D}K}(\tilde{p})$, $\phi_{\bar{D}^*K}(\tilde{p})$, $\phi_{\bar{D}K^*}(\tilde{p})$, $\phi_{\bar{D}^*K^*(^1P_1)}(\tilde{p})$, $\phi_{\bar{D}^*K^*(^3P_1)}(\tilde{p})$ and $\phi_{\bar{D}^*K^*(^5P_1)}(\tilde{p})$ are depicted in Fig. \ref{fig:wfp}. It is worth noticing that these two $P$-wave states are resonances relative to all scattering channels. The pole properties of two $P$-wave $T_{cs1-}(2900)$ and $T^{\prime}_{cs1-}(2900)$ in the coupled-channel analysis are listed in Table \ref{tab:x1} and \ref{tab:x1prime}, respectively. The dominant components of $T_{cs1-}(2900)$ and $T^{\prime}_{cs1-}(2900)$ are $\bar{D}^*K^*(^1P_1)$ and $\bar{D}^*K^*(^3P_1)$, respectively. Because the interaction matrix elements of the scattering $\bar{D}K^* \to \bar{D}^*K^*(^3P_1/^5P_1)$ and $\bar{D}^*K \to \bar{D}^*K^*(^3P_1/^5P_1)$ are not zero, the $\bar{D}K^*$ and $\bar{D}^*K$ channels bridge an indirect coupling between the $\bar{D}^*K^*(^3P_1)$ and $\bar{D}^*K^*(^5P_1)$ channels, which renders the $T^{\prime}_{cs1-}(2900)$ state no longer corresponding to a pure $^3P_1$ constituent of $\bar{D}^*K^*$. Now the $T^{\prime}_{cs1-}(2900)$ contains small $^1P_1$ and $^5P_1$ components. Additionally, the pole properties of these two $P$-wave resonances merely change with varying parameter $\lambda^{\prime}$ from $0.28\sim0.84$. This result further shows the reliability of our dynamical prediction of the existence of the $P$-wave $\bar{D}^*K^*$ resonances with $J^P=1^-$.

The direct interaction between the $\bar{D}^*K^*(^3P_1)$ and $\bar{D}K(^1P_1)$ channel is missing due to the similar reason with the analysis in Eq. (\ref{eq18}). Namely, the potential operators in Eqs. (\ref{eq28})-(\ref{eq29}) are symmetric under the particle exchange. Note that the  $X_1(2900)$ was observed in the final state of $\bar{D}K$. Although there exists a double pole structure of $X_1(2900)$, the $P$-wave $T_{cs1-}(2900)$ resonance makes a major contribution to the formation of the pronounced enhancement structure. From Table \ref{tab:x1prime}, we recommend searching for the dominant signal of $T^{\prime}_{cs1-}(2900)$ in the invariant mass spectrum of $\bar{D}K^*$ or $\bar{D}K\pi$.

\section{Predictions of more charmed-strange dimeson-type tetraquark states} \label{sec4}

\begin{figure*}[htb]
\begin{centering}
    \scalebox{1.0}{\includegraphics[width=\linewidth]{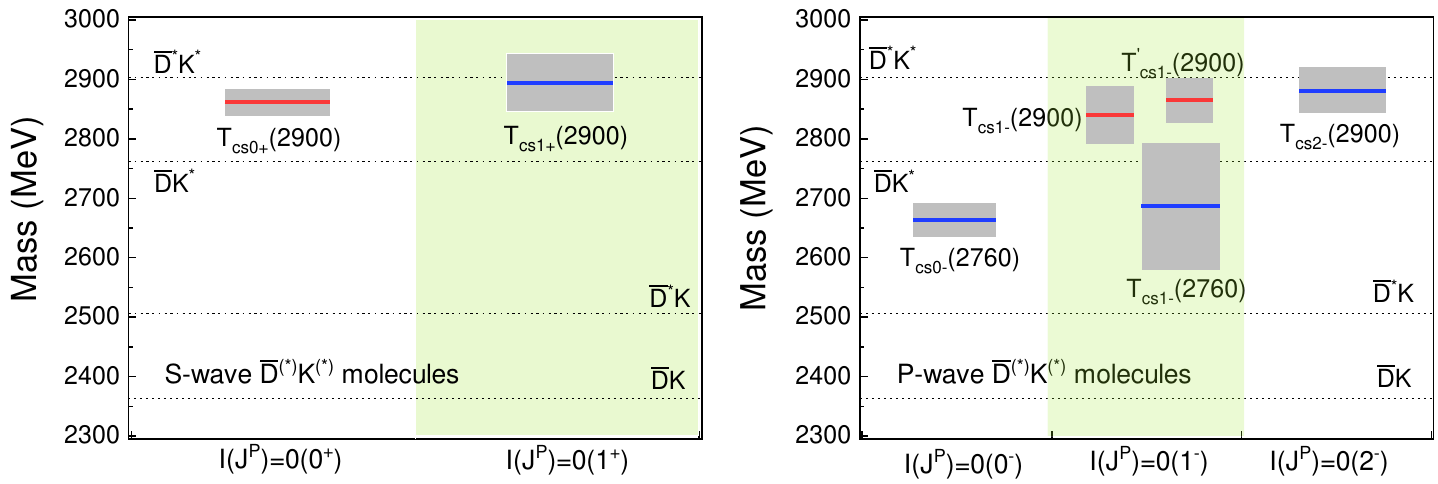}}
    \caption{ The spectrum of the charmed-strange dimeson-type tetraquarks. The horizontal and vertical coordinates represent the $I(J^P)$ quantum number and the masses, respectively. The color line and gray band correspond to the real part and the twice imaginary part of the energy pole, respectively. \label{fig:spec}}
\end{centering}
\end{figure*}

For the isoscalar $\bar{D}^{(*)}K^{(*)}$ scattering, there are some other $S$-wave channels with $J=1$ and $J=2$ and $P$-wave channels with $J=0$, $J=2$ and $J=3$. Based on the nice explanations of the experimentally observed $X_0(2900)$ and $X_1(2900)$ with three dimeson poles $T_{cs0+}(2900)$, $T_{cs1-}(2900)$ and $T^{\prime}_{cs1-}(2900)$, we also explore the potential poles in these channels and the relevant results are summarized in Table \ref{tab:morestates}. The $S$-wave channels with $J^P=1^+$ correspond to the three-channel coupling involving $\bar{D}^*K(^3S_1)$, $\bar{D}K^*(^3S_1)$ and $\bar{D}^*K^*(^3S_1)$.  We find a quasi-bound state pole $T_{cs1+}(2900)$ relative to the $\bar{D}^*K^*$ threshold. Its absolutely dominant component is the $\bar{D}^*K^*$, which is consistent with the conclusions in Refs. \cite{Liu:2020nil,Molina:2010tx,He:2020btl,Hu:2020mxp,Wang:2023hpp}. The pole position of $T_{cs1+}(2900)$ is $2.895-0.049i$ GeV, whose width is about two times larger than that of $T_{cs0+}(2900)$. We do not find any other resonance or bound state poles in the $1^+$ channels. We notice that a recent $S$-wave pole analysis for the $\bar{D}^*K$ and $\bar{D}K^*$ coupled-channel system suggested the existence of a resonance around 2.7 GeV \cite{Shi:2024squ}. However, it is worth mentioning that our conclusion is consistent with the dynamical quark model calculations with the complex scaling method in Ref. \cite{Chen:2023syh}. Finally, the $S$-wave partial-wave potential of $\bar{D}^*K^*$ with $J=2$ is repulsive at the large momentum, which cannot induce an observable pole.

We also find a new lower $P$-wave resonance $T_{cs1-}(2760)$ with a pole position of $2.686-0.107i$ GeV relative to the $\bar{D}K^*$ threshold, which has the same quantum number with $T_{cs1-}(2900)$ and $T^{\prime}_{cs1-}(2900)$. This pole locates on the Riemann sheet $(-,-,-,+)$, where the notations "$+$" and "$-$" represent for the physical and unphysical sheets, respectively, with respect to the four channels $\bar{D}K$, $\bar{D}^*K$, $\bar{D}K^*$ and $\bar{D}^*K^*$ successively. One can see that the pole width of $T_{cs1-}(2760)$ is too large to be easily detected.   The $P$-wave channel with $J^P=0^-$ also corresponds to a three-channel coupling including $\bar{D}^*K(^3P_0)$, $\bar{D}K^*(^3P_0)$ and $\bar{D}^*K^*(^3P_0)$. We obtain an energy pole with $2.664-0.029i$ GeV in the Riemann sheet of $(-,-,+)$. Additionally, we predict a $P$-wave charmed-strange near-threshold resonance $T_{cs2-}(2900)$ with $J=2$, which is absolutely governed by the $\bar{D}^*K^*(^5P_2)$ channel and the corresponding pole position is predicted to be $2.882-0.039i$ GeV.
We suggest experimental search for the $T_{cs0-}(2760)$ and $T_{cs2-}(2900)$ states by reconstructing the final states of $\bar{D}^*K$ in $B \to \bar{D}^{(*)}D^{(*)}K$ and $\bar{D}K\pi$ in $B \to \bar{D}^{(*)}D^{(*)}K\pi$. 
The $P$-wave system with $J=3$ corresponds to a single channel of $\bar{D}^*K^*(^5P_3)$, whose partial-wave potential is fully repulsive in total and so no dimeson pole exists.

\begin{table*}[htb]
\renewcommand{\arraystretch}{1.4}
\caption{The pole properties of four charmed-strange dimeson-type tetraquark states, which include an $S$-wave state $T_{cs1+}(2900)$, and three $P$-wave states $T_{cs1-}(2760)$, $T_{cs0-}(2760)$ and $T_{cs2-}(2900)$. The notation $\times, +, -$ in the fourth row stands for the forbidden, physical and unphysical Riemann sheet of the channel, respectively, and the four channels are $\bar{D}K$, $\bar{D}^*K$, $\bar{D}K^*$ and $\bar{D}^*K^*$ successively. \label{tab:morestates}}
\setlength{\tabcolsep}{8.0mm}
{
\centering
\begin{tabular}{ccccccccc}
\toprule[0.8pt]
\toprule[0.8pt]
     States & $T_{cs1+}(2900)$ & $T_{cs1-}(2760)$  & $T_{cs0-}(2760)$ & $T_{cs2-}(2900)$  \\
     \hline   
     $J^{P}$ & $1^+$ &$1^-$  & $0^-$ & $2^-$ \\
     \hline
      Pole (GeV)   & 2.895-0.049i  & 2.686-0.107i  &2.664-0.029i & 2.882-0.039i  \\
      \hline
       Sheet & ($\times$,$-$,$-$,$+$)  & ($-$,$-$,$-$,$+$)  &($\times$,$-$,$-$,+) & ($\times$,$-$,$-$,$-$)  \\
      \hline
      $\mathcal{P}(\bar{D}K)$    & \dots& (0.4+0.1i) \% & \dots& \dots  \\
       \hline
        $\mathcal{P}(\bar{D}^*K(^3S_1))$  & (-2.1-1.3i) \%  & \dots & \dots &  \dots   \\
       \hline
        $\mathcal{P}(\bar{D}^*K(^3P_0))$  & \dots  & \dots & (35.8+40.7i) \% &  \dots  \\
       \hline
        $\mathcal{P}(\bar{D}^*K(^3P_1))$  & \dots  & (-3.4-1.1i) \% & \dots &  \dots  \\
       \hline
      $\mathcal{P}(\bar{D}^*K(^3P_2))$ & \dots   & \dots &  \dots &  (0.2+0.2i) \%  \\
      \hline
       $\mathcal{P}(\bar{D}K^*(^3S_1))$  & (-7.7-6.8i) \%  & \dots &\dots & \dots  \\
       \hline
       $\mathcal{P}(\bar{D}K^*(^3P_0))$  & \dots  & \dots &(77.9-49.7i) \% & \dots  \\
       \hline
        $\mathcal{P}(\bar{D}K^*(^3P_1))$ & \dots   & (116.5+2.7i) \% &\dots& \dots   \\
       \hline
        $\mathcal{P}(\bar{D}K^*(^3P_2))$  & \dots  & \dots &\dots & (1.1+0.0i) \%   \\
         \hline
       $\mathcal{P}(\bar{D}^*K^*(^3S_1))$  & (109.8+8.1i) \%  & \dots &\dots & \dots   \\
       \hline
       $\mathcal{P}(\bar{D}^*K^*(^3P_0))$  & \dots  & \dots &(-13.7+9.0i) \% & \dots   \\
       \hline
       $\mathcal{P}(\bar{D}^*K^*(^1P_1))$  & \dots  & (-0.1+0.0i) \% & \dots & \dots   \\
        \hline
        $\mathcal{P}(\bar{D}^*K^*(^3P_1))$  & \dots  & (-2.4-1.1i) \% &\dots &  \dots  \\
       \hline
        $\mathcal{P}(\bar{D}^*K^*(^5P_1))$   & \dots & (-11.0-0.6i) \% &\dots & \dots   \\
       \hline
       $\mathcal{P}(\bar{D}^*K^*(^3P_2))$   & \dots & \dots &\dots & (-0.0-0.0i)\%   \\
       \hline
       $\mathcal{P}(\bar{D}^*K^*(^5P_2))$   & \dots & \dots &\dots & (98.7-0.2i)\%   \\
       \hline
         $Res(\bar{D}^*K)$  & 5.60   & \dots & \dots &  \dots  \\
       \hline
         $Res(\bar{D}K^*)$  &9.71  & 12.2 & 11.4 &  \dots  \\
          \hline
         $Res(\bar{D}^*K^*(^3S_1))$  &87.96  & \dots & \dots & \dots   \\
          \hline
         $Res(\bar{D}^*K^*(^3P_2))$  &\dots  & \dots & \dots & 0.0003    \\
        \hline
         $Res(\bar{D}^*K^*(^5P_2))$  &\dots  & \dots & \dots & 14.7   \\
      \bottomrule[0.8pt]
\bottomrule[0.8pt]
\end{tabular}
}
\end{table*}

\section{Summary and outlook} \label{sec5}

We have proposed a unified dimeson state scenario to explain the recently observed manifestly exotic charmed-strange tetraquark states $X_0(2900)$ and $X_1(2900)$ in the meson exchange model. Very intriguingly, we find the double pole structures of $X_1(2900)$, including two near-threshold $P$-wave $\bar{D}^*K^*$ resonances $T_{cs1-}(2900)$ and $T^{\prime}_{cs1-}(2900)$ on the unphysical Riemann sheet relative to the $\bar{D}^*K^*$ channel. The existence of these two $P$-wave resonances can be firmly established because their interactions are fully dominated by the well-known long-distance pion exchange and insensitive to the cutoff parameter in the regulator. With the coupled-channel analysis involving $\bar{D}K$, $\bar{D}^*K$, $\bar{D}K^*$ and $\bar{D}^*K^*$ and the energy-dependent width of the $K^*$ meson, 
both the experimental masses and widths of $X_0(2900)$ and $X_1(2900)$ remarkably coincide with our theoretical predictions of the $S$-wave pole $T_{cs0+}(2900)$ and two $P$-wave poles $T_{cs1-}(2900)$ and $T^{\prime}_{cs1-}(2900)$, respectively. Therefore, our research provides strong support for the dimeson state explanation of $X_0(2900)$ and $X_1(2900)$. 

In addition, we have predicted the existence of more isoscalar charmed-strange dimeson-type tetraquarks, which include an $S$-wave state $T_{cs1+}(2900)$ with quantum number $J^P=1^+$ in the Riemann sheet of $(\times,-,-,+)$, three $P$-wave states $T_{cs1-}(2760)$ with $J^P=1^-$ in the Riemann sheet $(-,-,-,+)$, $T_{cs0-}(2760)$ with $J^P=0^-$ in the Riemann sheet $(\times,-,-,+)$ and $T_{cs2-}(2900)$ with $J^P=2^-$ in the Riemann sheet $(\times,-,-,-)$, in which the notation $\times, +, -$ stands for the forbidden, physical and unphysical Riemann sheet of the channel, respectively, and the four channels are $\bar{D}K$, $\bar{D}^*K$, $\bar{D}K^*$ and $\bar{D}^*K^*$ successively. We suggest our experimental colleagues to search for these exotic tetraquark states and help construct the whole spectrum of the charmed-strange dimeson-type tetraquarks, which has been presented in Fig. \ref{fig:spec}. This should be accessible in future LHCb, Belle II and BESIII experiments.

The LHCb Collaboration recently observed the $T_{c\bar{s}0}^{a}(2900)$ state with $I(J^P)=1(0^+)$ \cite{LHCb:2022sfr,LHCb:2022lzp}, which is a good candidate of the $S$-wave $D^*K^*$ molecule state. Inspired by the double pole structures of $X_1(2900)$, we shall also expect the promising $P$-wave resonances in the $D^{(*)}K^{(*)}$ and $D_s^{(*)}\omega/D_s^{(*)}\rho$ system, and these investigations are ongoing.

\section*{Acknowledgement}

The authors thank Yan-Ke Chen for helpful discussions.
This project was supported by the National Natural Science Foundation of China under Grants No. 11975033, No. 12147168, No. 12070131001 and No. 12105072. This project was also funded by the Deutsche Forschungsgemeinschaft (DFG,
German Research Foundation, Project ID 196253076-TRR 110). J. Z. Wang is also supported by the National Postdoctoral Program for Innovative Talent. B. Wang is also supported by the Start-up Funds for Young Talents of Hebei University (No. 521100221021).

\appendix

\section{The partial-wave expansion potentials with the operators $\mathcal{O}_{1\sim6}$}

We list the partial-wave expansion  for the potentials $V_i=\mathcal{O}_{i}(p,p^{\prime},z) D(p,p^{\prime},z)$ $(i=1\sim6)$ under $S$-pave and $P$-wave, in which $z=\overset{\rightarrow}{p}\cdot\overset{\rightarrow}{p}^{\prime}/pp^{\prime}$ and $D(p,p^{\prime},z)=1/(p^2+p^{\prime2}-2pp^{\prime}z+m^2)$. For the operator $\mathcal{O}_1$, there are
\begin{eqnarray}
&&V_{S}^{J=0/1/2}=2\pi\int_{-1}^{1}D(p,p^{\prime},z)dz, \nonumber \\
&&V_{P}^{J=0/1/2/3}=2\pi\int_{-1}^{1}D(p,p^{\prime},z)zdz.  
\end{eqnarray}
For the operator $\mathcal{O}_2$, there are
\begin{eqnarray}
&&V_{S}^{J=0}=\frac{4\pi}{3}\int_{-1}^{1}D(p,p^{\prime},z)(p^2+p^{\prime2}-2pp^{\prime}z)dz, \nonumber \\
&&V_{S}^{J=1}=\frac{2\pi}{3}\int_{-1}^{1}D(p,p^{\prime},z)(p^2+p^{\prime2}-2pp^{\prime}z)dz, \nonumber \\
&&V_{S}^{J=2}=-\frac{2\pi}{3}\int_{-1}^{1}D(p,p^{\prime},z)(p^2+p^{\prime2}-2pp^{\prime}z)dz, \nonumber \\
&&V_{P}^{J=0}=2\pi\int_{-1}^{1}D(p,p^{\prime},z)(p^2z+p^{\prime2}z-pp^{\prime}(1+z^2))dz, \nonumber \\
&&V_{P}^{J=1,S=0}=\frac{4\pi}{3}\int_{-1}^{1}D(p,p^{\prime},z)(p^2+p^{\prime2}-pp^{\prime}z)zdz, \nonumber \\
&&V_{P}^{J=1,S=1}=\pi\int_{-1}^{1}D(p,p^{\prime},z)(pp^{\prime})(1-z^2)dz, \nonumber \\
&&V_{P}^{J=1,S=2}=\frac{\pi}{15}\int_{-1}^{1}D(p,p^{\prime},z)\nonumber \\
&&~~~~~~~~~~~~~~~~~\times(4p^2z+4p^{\prime2}z-pp^{\prime}(21-13z^2))dz, \nonumber \\
&&V_{P}^{J=1,S=0\otimes S=2}=-\frac{2\pi}{3\sqrt{5}}\int_{-1}^{1}D(p,p^{\prime},z) \nonumber \\
&&~~~~~~~~~~~~~~~~~\times(2p^2z+2p^{\prime2}z-pp^{\prime}(3+z^2))dz, \nonumber \\
&&V_{P}^{J=2,S=1}=\frac{\pi}{5}\int_{-1}^{1}D(p,p^{\prime},z)\nonumber \\
&&~~~~~~~~~~~~~~~~~\times(4p^2z+4p^{\prime2}z-pp^{\prime}(1+7z^2))dz, \nonumber \\
&&V_{P}^{J=2,S=2}=-\frac{\pi}{5}\int_{-1}^{1}D(p,p^{\prime},z)\nonumber \\
&&~~~~~~~~~~~~~~~~~\times(8p^2z+8p^{\prime2}z-pp^{\prime}(7+9z^2))dz, \nonumber \\
&&V_{P}^{J=3,S=2}=-\frac{2\pi}{5}\int_{-1}^{1}D(p,p^{\prime},z)\nonumber \\
&&~~~~~~~~~~~~~~~~~\times(p^2z+p^{\prime2}z+pp^{\prime}(1-3z^2))dz. 
\end{eqnarray}
For the operator $\mathcal{O}_3$, there are
\begin{eqnarray} &&V_{S}^{J=0}=2\pi\int_{-1}^{1}D(p,p^{\prime},z)(p^2+p^{\prime2}-2pp^{\prime}z)dz, \nonumber \\
&&V_{P}^{J=1,S=0}=2\pi\int_{-1}^{1}D(p,p^{\prime},z)(p^2+p^{\prime2}-pp^{\prime}z)zdz, \nonumber \\
&&V_{P}^{J=1,S=0\otimes S=2}=\frac{2\pi}{\sqrt{5}}\int_{-1}^{1}D(p,p^{\prime},z) \nonumber \\
&&~~~~~~~~~~~~~~~~~\times(2p^2z+2p^{\prime2}z-pp^{\prime}(3+z^2))dz. 
\end{eqnarray}
For the operator $\mathcal{O}_4$, there are
\begin{eqnarray} &&V_{S}^{J=0}=\frac{2\pi}{3}\int_{-1}^{1}D(p,p^{\prime},z)(p^2+p^{\prime2}-2pp^{\prime}z)dz, \nonumber \\
&&V_{S}^{J=1}=-\frac{2\pi}{3}\int_{-1}^{1}D(p,p^{\prime},z)(p^2+p^{\prime2}-2pp^{\prime}z)dz, \nonumber \\
&&V_{S}^{J=2}=\frac{2\pi}{3}\int_{-1}^{1}D(p,p^{\prime},z)(p^2+p^{\prime2}-2pp^{\prime}z)dz, \nonumber \\
&&V_{P}^{J=1,S=0}=\frac{2\pi}{3}\int_{-1}^{1}D(p,p^{\prime},z)(p^2+p^{\prime2}-pp^{\prime}z)zdz, \nonumber \\
&&V_{P}^{J=1,S=1}=-\frac{\pi}{5}\int_{-1}^{1}D(p,p^{\prime},z)\nonumber \\&&~~~~~~~~~~~~~~~~~\times(5p^2z+5p^{\prime2}z-pp^{\prime}(5+3z))dz, \nonumber \\
&&V_{P}^{J=1,S=2}=\frac{\pi}{30}\int_{-1}^{1}D(p,p^{\prime},z)\nonumber \\&&~~~~~~~~~~~~~~~~~\times(34p^2z+34p^{\prime2}z-pp^{\prime}(21+47z))dz, \nonumber \\
&&V_{P}^{J=1,S=0\otimes S=2}=\frac{2\pi}{3\sqrt{5}}\int_{-1}^{1}D(p,p^{\prime},z) \nonumber \\
&&~~~~~~~~~~~~~~~~~\times(2p^2z+2p^{\prime2}z-pp^{\prime}(3+z^2))dz, \nonumber\\
&&V_{P}^{J=1,S=1\otimes S=2}=\frac{\sqrt{3}\pi}{2\sqrt{5}}\int_{-1}^{1}D(p,p^{\prime},z) \nonumber \\
&&~~~~~~~~~~~~~~~~~\times(2p^2z+2p^{\prime2}z-pp^{\prime}(3+z^2))dz, \nonumber \\
&&V_{P}^{J=2,S=1}=-\frac{\pi}{10}\int_{-1}^{1}D(p,p^{\prime},z)\nonumber \\&&~~~~~~~~~~~~~~~~~\times(6p^2z+6p^{\prime2}z+pp^{\prime}(1-13z^2))dz, \nonumber \\
&&V_{P}^{J=2,S=2}=\frac{\pi}{10}\int_{-1}^{1}D(p,p^{\prime},z)\nonumber \\&&~~~~~~~~~~~~~~~~~\times(2p^2z+2p^{\prime2}z+pp^{\prime}(7-11z^2))dz, \nonumber \\
&&V_{P}^{J=2,S=1\otimes S=2}=-\frac{\sqrt{3}\pi}{10}\int_{-1}^{1}D(p,p^{\prime},z) \nonumber \\
&&~~~~~~~~~~~~~~~~~\times(2p^2z+2p^{\prime2}z-pp^{\prime}(3+z^2))dz, \nonumber \\
&&V_{P}^{J=3,S=2}=\frac{\pi}{5}\int_{-1}^{1}D(p,p^{\prime},z)\nonumber \\&&~~~~~~~~~~~~~~~~~\times(4p^2z+4p^{\prime2}z-pp^{\prime}(1+7z^2))dz.
\end{eqnarray}
For the operator $\mathcal{O}_5$, there are
\begin{eqnarray} &&V_{S}^{J=0}=\frac{2\pi}{3}\int_{-1}^{1}D(p,p^{\prime},z)(p^2+p^{\prime2}-2pp^{\prime}z)dz, \nonumber \\
&&V_{S}^{J=1}=-\frac{2\pi}{3}\int_{-1}^{1}D(p,p^{\prime},z)(p^2+p^{\prime2}-2pp^{\prime}z)dz, \nonumber \\
&&V_{S}^{J=2}=\frac{2\pi}{3}\int_{-1}^{1}D(p,p^{\prime},z)(p^2+p^{\prime2}-2pp^{\prime}z)dz, \nonumber \\
&&V_{P}^{J=1,S=0}=\frac{2\pi}{3}\int_{-1}^{1}D(p,p^{\prime},z)(p^2+p^{\prime2}-pp^{\prime}z)zdz, \nonumber \\
&&V_{P}^{J=1,S=1}=-\frac{\pi}{5}\int_{-1}^{1}D(p,p^{\prime},z)\nonumber \\&&~~~~~~~~~~~~~~~~~\times(5p^2z+5p^{\prime2}z-pp^{\prime}(5+3z))dz, \nonumber \\
&&V_{P}^{J=1,S=2}=\frac{\pi}{30}\int_{-1}^{1}D(p,p^{\prime},z)\nonumber \\&&~~~~~~~~~~~~~~~~~\times(34p^2z+34p^{\prime2}z-pp^{\prime}(21+47z))dz, \nonumber \\
&&V_{P}^{J=1,S=0\otimes S=2}=\frac{2\pi}{3\sqrt{5}}\int_{-1}^{1}D(p,p^{\prime},z) \nonumber \\
&&~~~~~~~~~~~~~~~~~\times(2p^2z+2p^{\prime2}z-pp^{\prime}(3+z^2))dz, \nonumber\\
&&V_{P}^{J=1,S=1\otimes S=2}=-\frac{\sqrt{3}\pi}{2\sqrt{5}}\int_{-1}^{1}D(p,p^{\prime},z) \nonumber \\
&&~~~~~~~~~~~~~~~~~\times(2p^2z+2p^{\prime2}z-pp^{\prime}(3+z^2))dz, \nonumber \\
&&V_{P}^{J=2,S=1}=-\frac{\pi}{10}\int_{-1}^{1}D(p,p^{\prime},z)\nonumber \\&&~~~~~~~~~~~~~~~~~\times(6p^2z+6p^{\prime2}z+pp^{\prime}(1-13z^2))dz, \nonumber \\
&&V_{P}^{J=2,S=2}=\frac{\pi}{10}\int_{-1}^{1}D(p,p^{\prime},z)\nonumber \\&&~~~~~~~~~~~~~~~~~\times(2p^2z+2p^{\prime2}z+pp^{\prime}(7-11z^2))dz, \nonumber \\
&&V_{P}^{J=2,S=1\otimes S=2}=\frac{\sqrt{3}\pi}{10}\int_{-1}^{1}D(p,p^{\prime},z) \nonumber \\
&&~~~~~~~~~~~~~~~~~\times(2p^2z+2p^{\prime2}z-pp^{\prime}(3+z^2))dz, \nonumber \\
&&V_{P}^{J=3,S=2}=\frac{\pi}{5}\int_{-1}^{1}D(p,p^{\prime},z)\nonumber \\&&~~~~~~~~~~~~~~~~~\times(4p^2z+4p^{\prime2}z-pp^{\prime}(1+7z^2))dz.
\end{eqnarray}
For the operator $\mathcal{O}_6$, there are
\begin{eqnarray} &&V_{S}^{J=0}=2\pi\int_{-1}^{1}D(p,p^{\prime},z)(p^2+p^{\prime2}-2pp^{\prime}z)dz, \nonumber \\
&&V_{P}^{J=1,S=0}=2\pi\int_{-1}^{1}D(p,p^{\prime},z)(p^2+p^{\prime2}-pp^{\prime}z)zdz, \nonumber \\
&&V_{P}^{J=1,S=2\otimes S=0}=\frac{2\pi}{\sqrt{5}}\int_{-1}^{1}D(p,p^{\prime},z) \nonumber \\
&&~~~~~~~~~~~~~~~~~\times(2p^2z+2p^{\prime2}z-pp^{\prime}(3+z^2))dz. 
\end{eqnarray}

\bibliography{refs}

\begin{thebibliography}{61}%
\makeatletter
\providecommand \@ifxundefined [1]{%
 \@ifx{#1\undefined}
}%
\providecommand \@ifnum [1]{%
 \ifnum #1\expandafter \@firstoftwo
 \else \expandafter \@secondoftwo
 \fi
}%
\providecommand \@ifx [1]{%
 \ifx #1\expandafter \@firstoftwo
 \else \expandafter \@secondoftwo
 \fi
}%
\providecommand \natexlab [1]{#1}%
\providecommand \enquote  [1]{``#1''}%
\providecommand \bibnamefont  [1]{#1}%
\providecommand \bibfnamefont [1]{#1}%
\providecommand \citenamefont [1]{#1}%
\providecommand \href@noop [0]{\@secondoftwo}%
\providecommand \href [0]{\begingroup \@sanitize@url \@href}%
\providecommand \@href[1]{\@@startlink{#1}\@@href}%
\providecommand \@@href[1]{\endgroup#1\@@endlink}%
\providecommand \@sanitize@url [0]{\catcode `\\12\catcode `\$12\catcode `\&12\catcode `\#12\catcode `\^12\catcode `\_12\catcode `\%12\relax}%
\providecommand \@@startlink[1]{}%
\providecommand \@@endlink[0]{}%
\providecommand \url  [0]{\begingroup\@sanitize@url \@url }%
\providecommand \@url [1]{\endgroup\@href {#1}{\urlprefix }}%
\providecommand \urlprefix  [0]{URL }%
\providecommand \Eprint [0]{\href }%
\providecommand \doibase [0]{http://dx.doi.org/}%
\providecommand \selectlanguage [0]{\@gobble}%
\providecommand \bibinfo  [0]{\@secondoftwo}%
\providecommand \bibfield  [0]{\@secondoftwo}%
\providecommand \translation [1]{[#1]}%
\providecommand \BibitemOpen [0]{}%
\providecommand \bibitemStop [0]{}%
\providecommand \bibitemNoStop [0]{.\EOS\space}%
\providecommand \EOS [0]{\spacefactor3000\relax}%
\providecommand \BibitemShut  [1]{\csname bibitem#1\endcsname}%
\let\auto@bib@innerbib\@empty
\bibitem [{\citenamefont {Chen}\ \emph {et~al.}(2016)\citenamefont {Chen}, \citenamefont {Chen}, \citenamefont {Liu},\ and\ \citenamefont {Zhu}}]{Chen:2016qju}%
  \BibitemOpen
  \bibfield  {author} {\bibinfo {author} {\bibfnamefont {H.-X.}\ \bibnamefont {Chen}}, \bibinfo {author} {\bibfnamefont {W.}~\bibnamefont {Chen}}, \bibinfo {author} {\bibfnamefont {X.}~\bibnamefont {Liu}}, \ and\ \bibinfo {author} {\bibfnamefont {S.-L.}\ \bibnamefont {Zhu}},\ }\href {\doibase 10.1016/j.physrep.2016.05.004} {\bibfield  {journal} {\bibinfo  {journal} {Phys. Rept.}\ }\textbf {\bibinfo {volume} {639}},\ \bibinfo {pages} {1} (\bibinfo {year} {2016})},\ \Eprint {http://arxiv.org/abs/1601.02092} {arXiv:1601.02092 [hep-ph]} \BibitemShut {NoStop}%
\bibitem [{\citenamefont {Guo}\ \emph {et~al.}(2018)\citenamefont {Guo}, \citenamefont {Hanhart}, \citenamefont {Mei\ss{}ner}, \citenamefont {Wang}, \citenamefont {Zhao},\ and\ \citenamefont {Zou}}]{Guo:2017jvc}%
  \BibitemOpen
  \bibfield  {author} {\bibinfo {author} {\bibfnamefont {F.-K.}\ \bibnamefont {Guo}}, \bibinfo {author} {\bibfnamefont {C.}~\bibnamefont {Hanhart}}, \bibinfo {author} {\bibfnamefont {U.-G.}\ \bibnamefont {Mei\ss{}ner}}, \bibinfo {author} {\bibfnamefont {Q.}~\bibnamefont {Wang}}, \bibinfo {author} {\bibfnamefont {Q.}~\bibnamefont {Zhao}}, \ and\ \bibinfo {author} {\bibfnamefont {B.-S.}\ \bibnamefont {Zou}},\ }\href {\doibase 10.1103/RevModPhys.90.015004} {\bibfield  {journal} {\bibinfo  {journal} {Rev. Mod. Phys.}\ }\textbf {\bibinfo {volume} {90}},\ \bibinfo {pages} {015004} (\bibinfo {year} {2018})},\ \Eprint {http://arxiv.org/abs/1705.00141} {arXiv:1705.00141 [hep-ph]} \BibitemShut {NoStop}%
\bibitem [{\citenamefont {Liu}\ \emph {et~al.}(2019)\citenamefont {Liu}, \citenamefont {Chen}, \citenamefont {Chen}, \citenamefont {Liu},\ and\ \citenamefont {Zhu}}]{Liu:2019zoy}%
  \BibitemOpen
  \bibfield  {author} {\bibinfo {author} {\bibfnamefont {Y.-R.}\ \bibnamefont {Liu}}, \bibinfo {author} {\bibfnamefont {H.-X.}\ \bibnamefont {Chen}}, \bibinfo {author} {\bibfnamefont {W.}~\bibnamefont {Chen}}, \bibinfo {author} {\bibfnamefont {X.}~\bibnamefont {Liu}}, \ and\ \bibinfo {author} {\bibfnamefont {S.-L.}\ \bibnamefont {Zhu}},\ }\href {\doibase 10.1016/j.ppnp.2019.04.003} {\bibfield  {journal} {\bibinfo  {journal} {Prog. Part. Nucl. Phys.}\ }\textbf {\bibinfo {volume} {107}},\ \bibinfo {pages} {237} (\bibinfo {year} {2019})},\ \Eprint {http://arxiv.org/abs/1903.11976} {arXiv:1903.11976 [hep-ph]} \BibitemShut {NoStop}%
\bibitem [{\citenamefont {Lebed}\ \emph {et~al.}(2017)\citenamefont {Lebed}, \citenamefont {Mitchell},\ and\ \citenamefont {Swanson}}]{Lebed:2016hpi}%
  \BibitemOpen
  \bibfield  {author} {\bibinfo {author} {\bibfnamefont {R.~F.}\ \bibnamefont {Lebed}}, \bibinfo {author} {\bibfnamefont {R.~E.}\ \bibnamefont {Mitchell}}, \ and\ \bibinfo {author} {\bibfnamefont {E.~S.}\ \bibnamefont {Swanson}},\ }\href {\doibase 10.1016/j.ppnp.2016.11.003} {\bibfield  {journal} {\bibinfo  {journal} {Prog. Part. Nucl. Phys.}\ }\textbf {\bibinfo {volume} {93}},\ \bibinfo {pages} {143} (\bibinfo {year} {2017})},\ \Eprint {http://arxiv.org/abs/1610.04528} {arXiv:1610.04528 [hep-ph]} \BibitemShut {NoStop}%
\bibitem [{\citenamefont {Esposito}\ \emph {et~al.}(2017)\citenamefont {Esposito}, \citenamefont {Pilloni},\ and\ \citenamefont {Polosa}}]{Esposito:2016noz}%
  \BibitemOpen
  \bibfield  {author} {\bibinfo {author} {\bibfnamefont {A.}~\bibnamefont {Esposito}}, \bibinfo {author} {\bibfnamefont {A.}~\bibnamefont {Pilloni}}, \ and\ \bibinfo {author} {\bibfnamefont {A.~D.}\ \bibnamefont {Polosa}},\ }\href {\doibase 10.1016/j.physrep.2016.11.002} {\bibfield  {journal} {\bibinfo  {journal} {Phys. Rept.}\ }\textbf {\bibinfo {volume} {668}},\ \bibinfo {pages} {1} (\bibinfo {year} {2017})},\ \Eprint {http://arxiv.org/abs/1611.07920} {arXiv:1611.07920 [hep-ph]} \BibitemShut {NoStop}%
\bibitem [{\citenamefont {Brambilla}\ \emph {et~al.}(2020)\citenamefont {Brambilla}, \citenamefont {Eidelman}, \citenamefont {Hanhart}, \citenamefont {Nefediev}, \citenamefont {Shen}, \citenamefont {Thomas}, \citenamefont {Vairo},\ and\ \citenamefont {Yuan}}]{Brambilla:2019esw}%
  \BibitemOpen
  \bibfield  {author} {\bibinfo {author} {\bibfnamefont {N.}~\bibnamefont {Brambilla}}, \bibinfo {author} {\bibfnamefont {S.}~\bibnamefont {Eidelman}}, \bibinfo {author} {\bibfnamefont {C.}~\bibnamefont {Hanhart}}, \bibinfo {author} {\bibfnamefont {A.}~\bibnamefont {Nefediev}}, \bibinfo {author} {\bibfnamefont {C.-P.}\ \bibnamefont {Shen}}, \bibinfo {author} {\bibfnamefont {C.~E.}\ \bibnamefont {Thomas}}, \bibinfo {author} {\bibfnamefont {A.}~\bibnamefont {Vairo}}, \ and\ \bibinfo {author} {\bibfnamefont {C.-Z.}\ \bibnamefont {Yuan}},\ }\href {\doibase 10.1016/j.physrep.2020.05.001} {\bibfield  {journal} {\bibinfo  {journal} {Phys. Rept.}\ }\textbf {\bibinfo {volume} {873}},\ \bibinfo {pages} {1} (\bibinfo {year} {2020})},\ \Eprint {http://arxiv.org/abs/1907.07583} {arXiv:1907.07583 [hep-ex]} \BibitemShut {NoStop}%
\bibitem [{\citenamefont {Chen}\ \emph {et~al.}(2023{\natexlab{a}})\citenamefont {Chen}, \citenamefont {Li}, \citenamefont {Qian}, \citenamefont {Shen}, \citenamefont {Xie}, \citenamefont {Yang}, \citenamefont {Zhang},\ and\ \citenamefont {Zhang}}]{Chen:2021ftn}%
  \BibitemOpen
  \bibfield  {author} {\bibinfo {author} {\bibfnamefont {S.}~\bibnamefont {Chen}}, \bibinfo {author} {\bibfnamefont {Y.}~\bibnamefont {Li}}, \bibinfo {author} {\bibfnamefont {W.}~\bibnamefont {Qian}}, \bibinfo {author} {\bibfnamefont {Z.}~\bibnamefont {Shen}}, \bibinfo {author} {\bibfnamefont {Y.}~\bibnamefont {Xie}}, \bibinfo {author} {\bibfnamefont {Z.}~\bibnamefont {Yang}}, \bibinfo {author} {\bibfnamefont {L.}~\bibnamefont {Zhang}}, \ and\ \bibinfo {author} {\bibfnamefont {Y.}~\bibnamefont {Zhang}},\ }\href {\doibase 10.1007/s11467-022-1247-1} {\bibfield  {journal} {\bibinfo  {journal} {Front. Phys.}\ }\textbf {\bibinfo {volume} {18}},\ \bibinfo {pages} {44601} (\bibinfo {year} {2023}{\natexlab{a}})},\ \Eprint {http://arxiv.org/abs/2111.14360} {arXiv:2111.14360 [hep-ex]} \BibitemShut {NoStop}%
\bibitem [{\citenamefont {Chen}\ \emph {et~al.}(2023{\natexlab{b}})\citenamefont {Chen}, \citenamefont {Chen}, \citenamefont {Liu}, \citenamefont {Liu},\ and\ \citenamefont {Zhu}}]{Chen:2022asf}%
  \BibitemOpen
  \bibfield  {author} {\bibinfo {author} {\bibfnamefont {H.-X.}\ \bibnamefont {Chen}}, \bibinfo {author} {\bibfnamefont {W.}~\bibnamefont {Chen}}, \bibinfo {author} {\bibfnamefont {X.}~\bibnamefont {Liu}}, \bibinfo {author} {\bibfnamefont {Y.-R.}\ \bibnamefont {Liu}}, \ and\ \bibinfo {author} {\bibfnamefont {S.-L.}\ \bibnamefont {Zhu}},\ }\href {\doibase 10.1088/1361-6633/aca3b6} {\bibfield  {journal} {\bibinfo  {journal} {Rept. Prog. Phys.}\ }\textbf {\bibinfo {volume} {86}},\ \bibinfo {pages} {026201} (\bibinfo {year} {2023}{\natexlab{b}})},\ \Eprint {http://arxiv.org/abs/2204.02649} {arXiv:2204.02649 [hep-ph]} \BibitemShut {NoStop}%
\bibitem [{\citenamefont {Meng}\ \emph {et~al.}(2023)\citenamefont {Meng}, \citenamefont {Wang}, \citenamefont {Wang},\ and\ \citenamefont {Zhu}}]{Meng:2022ozq}%
  \BibitemOpen
  \bibfield  {author} {\bibinfo {author} {\bibfnamefont {L.}~\bibnamefont {Meng}}, \bibinfo {author} {\bibfnamefont {B.}~\bibnamefont {Wang}}, \bibinfo {author} {\bibfnamefont {G.-J.}\ \bibnamefont {Wang}}, \ and\ \bibinfo {author} {\bibfnamefont {S.-L.}\ \bibnamefont {Zhu}},\ }\href {\doibase 10.1016/j.physrep.2023.04.003} {\bibfield  {journal} {\bibinfo  {journal} {Phys. Rept.}\ }\textbf {\bibinfo {volume} {1019}},\ \bibinfo {pages} {1} (\bibinfo {year} {2023})},\ \Eprint {http://arxiv.org/abs/2204.08716} {arXiv:2204.08716 [hep-ph]} \BibitemShut {NoStop}%
\bibitem [{\citenamefont {Choi}\ \emph {et~al.}(2003)\citenamefont {Choi} \emph {et~al.}}]{Belle:2003nnu}%
  \BibitemOpen
  \bibfield  {author} {\bibinfo {author} {\bibfnamefont {S.~K.}\ \bibnamefont {Choi}} \emph {et~al.} (\bibinfo {collaboration} {Belle}),\ }\href {\doibase 10.1103/PhysRevLett.91.262001} {\bibfield  {journal} {\bibinfo  {journal} {Phys. Rev. Lett.}\ }\textbf {\bibinfo {volume} {91}},\ \bibinfo {pages} {262001} (\bibinfo {year} {2003})},\ \Eprint {http://arxiv.org/abs/hep-ex/0309032} {arXiv:hep-ex/0309032} \BibitemShut {NoStop}%
\bibitem [{\citenamefont {Ablikim}\ \emph {et~al.}(2013)\citenamefont {Ablikim} \emph {et~al.}}]{BESIII:2013ris}%
  \BibitemOpen
  \bibfield  {author} {\bibinfo {author} {\bibfnamefont {M.}~\bibnamefont {Ablikim}} \emph {et~al.} (\bibinfo {collaboration} {BESIII}),\ }\href {\doibase 10.1103/PhysRevLett.110.252001} {\bibfield  {journal} {\bibinfo  {journal} {Phys. Rev. Lett.}\ }\textbf {\bibinfo {volume} {110}},\ \bibinfo {pages} {252001} (\bibinfo {year} {2013})},\ \Eprint {http://arxiv.org/abs/1303.5949} {arXiv:1303.5949 [hep-ex]} \BibitemShut {NoStop}%
\bibitem [{\citenamefont {Aaij}\ \emph {et~al.}(2022{\natexlab{a}})\citenamefont {Aaij} \emph {et~al.}}]{LHCb:2021vvq}%
  \BibitemOpen
  \bibfield  {author} {\bibinfo {author} {\bibfnamefont {R.}~\bibnamefont {Aaij}} \emph {et~al.} (\bibinfo {collaboration} {LHCb}),\ }\href {\doibase 10.1038/s41567-022-01614-y} {\bibfield  {journal} {\bibinfo  {journal} {Nature Phys.}\ }\textbf {\bibinfo {volume} {18}},\ \bibinfo {pages} {751} (\bibinfo {year} {2022}{\natexlab{a}})},\ \Eprint {http://arxiv.org/abs/2109.01038} {arXiv:2109.01038 [hep-ex]} \BibitemShut {NoStop}%
\bibitem [{\citenamefont {Aaij}\ \emph {et~al.}(2022{\natexlab{b}})\citenamefont {Aaij} \emph {et~al.}}]{LHCb:2021auc}%
  \BibitemOpen
  \bibfield  {author} {\bibinfo {author} {\bibfnamefont {R.}~\bibnamefont {Aaij}} \emph {et~al.} (\bibinfo {collaboration} {LHCb}),\ }\href {\doibase 10.1038/s41467-022-30206-w} {\bibfield  {journal} {\bibinfo  {journal} {Nature Commun.}\ }\textbf {\bibinfo {volume} {13}},\ \bibinfo {pages} {3351} (\bibinfo {year} {2022}{\natexlab{b}})},\ \Eprint {http://arxiv.org/abs/2109.01056} {arXiv:2109.01056 [hep-ex]} \BibitemShut {NoStop}%
\bibitem [{\citenamefont {Workman}\ \emph {et~al.}(2022)\citenamefont {Workman} \emph {et~al.}}]{Workman:2022ynf}%
  \BibitemOpen
  \bibfield  {author} {\bibinfo {author} {\bibfnamefont {R.~L.}\ \bibnamefont {Workman}} \emph {et~al.} (\bibinfo {collaboration} {Particle Data Group}),\ }\href {\doibase 10.1093/ptep/ptac097} {\bibfield  {journal} {\bibinfo  {journal} {PTEP}\ }\textbf {\bibinfo {volume} {2022}},\ \bibinfo {pages} {083C01} (\bibinfo {year} {2022})}\BibitemShut {NoStop}%
\bibitem [{\citenamefont {Aaij}\ \emph {et~al.}(2019)\citenamefont {Aaij} \emph {et~al.}}]{LHCb:2019kea}%
  \BibitemOpen
  \bibfield  {author} {\bibinfo {author} {\bibfnamefont {R.}~\bibnamefont {Aaij}} \emph {et~al.} (\bibinfo {collaboration} {LHCb}),\ }\href {\doibase 10.1103/PhysRevLett.122.222001} {\bibfield  {journal} {\bibinfo  {journal} {Phys. Rev. Lett.}\ }\textbf {\bibinfo {volume} {122}},\ \bibinfo {pages} {222001} (\bibinfo {year} {2019})},\ \Eprint {http://arxiv.org/abs/1904.03947} {arXiv:1904.03947 [hep-ex]} \BibitemShut {NoStop}%
\bibitem [{\citenamefont {Gershon}(2022)}]{Gershon:2022xnn}%
  \BibitemOpen
  \bibfield  {author} {\bibinfo {author} {\bibfnamefont {T.}~\bibnamefont {Gershon}} (\bibinfo {collaboration} {LHCb}),\ }\href@noop {} {\  (\bibinfo {year} {2022})},\ \Eprint {http://arxiv.org/abs/2206.15233} {arXiv:2206.15233 [hep-ex]} \BibitemShut {NoStop}%
\bibitem [{\citenamefont {Aaij}\ \emph {et~al.}(2020{\natexlab{a}})\citenamefont {Aaij} \emph {et~al.}}]{LHCb:2020bls}%
  \BibitemOpen
  \bibfield  {author} {\bibinfo {author} {\bibfnamefont {R.}~\bibnamefont {Aaij}} \emph {et~al.} (\bibinfo {collaboration} {LHCb}),\ }\href {\doibase 10.1103/PhysRevLett.125.242001} {\bibfield  {journal} {\bibinfo  {journal} {Phys. Rev. Lett.}\ }\textbf {\bibinfo {volume} {125}},\ \bibinfo {pages} {242001} (\bibinfo {year} {2020}{\natexlab{a}})},\ \Eprint {http://arxiv.org/abs/2009.00025} {arXiv:2009.00025 [hep-ex]} \BibitemShut {NoStop}%
\bibitem [{\citenamefont {Aaij}\ \emph {et~al.}(2020{\natexlab{b}})\citenamefont {Aaij} \emph {et~al.}}]{LHCb:2020pxc}%
  \BibitemOpen
  \bibfield  {author} {\bibinfo {author} {\bibfnamefont {R.}~\bibnamefont {Aaij}} \emph {et~al.} (\bibinfo {collaboration} {LHCb}),\ }\href {\doibase 10.1103/PhysRevD.102.112003} {\bibfield  {journal} {\bibinfo  {journal} {Phys. Rev. D}\ }\textbf {\bibinfo {volume} {102}},\ \bibinfo {pages} {112003} (\bibinfo {year} {2020}{\natexlab{b}})},\ \Eprint {http://arxiv.org/abs/2009.00026} {arXiv:2009.00026 [hep-ex]} \BibitemShut {NoStop}%
\bibitem [{\citenamefont {Chen}\ \emph {et~al.}(2020)\citenamefont {Chen}, \citenamefont {Chen}, \citenamefont {Dong},\ and\ \citenamefont {Su}}]{Chen:2020aos}%
  \BibitemOpen
  \bibfield  {author} {\bibinfo {author} {\bibfnamefont {H.-X.}\ \bibnamefont {Chen}}, \bibinfo {author} {\bibfnamefont {W.}~\bibnamefont {Chen}}, \bibinfo {author} {\bibfnamefont {R.-R.}\ \bibnamefont {Dong}}, \ and\ \bibinfo {author} {\bibfnamefont {N.}~\bibnamefont {Su}},\ }\href {\doibase 10.1088/0256-307X/37/10/101201} {\bibfield  {journal} {\bibinfo  {journal} {Chin. Phys. Lett.}\ }\textbf {\bibinfo {volume} {37}},\ \bibinfo {pages} {101201} (\bibinfo {year} {2020})},\ \Eprint {http://arxiv.org/abs/2008.07516} {arXiv:2008.07516 [hep-ph]} \BibitemShut {NoStop}%
\bibitem [{\citenamefont {He}\ and\ \citenamefont {Chen}(2021)}]{He:2020btl}%
  \BibitemOpen
  \bibfield  {author} {\bibinfo {author} {\bibfnamefont {J.}~\bibnamefont {He}}\ and\ \bibinfo {author} {\bibfnamefont {D.-Y.}\ \bibnamefont {Chen}},\ }\href {\doibase 10.1088/1674-1137/abeda8} {\bibfield  {journal} {\bibinfo  {journal} {Chin. Phys. C}\ }\textbf {\bibinfo {volume} {45}},\ \bibinfo {pages} {063102} (\bibinfo {year} {2021})},\ \Eprint {http://arxiv.org/abs/2008.07782} {arXiv:2008.07782 [hep-ph]} \BibitemShut {NoStop}%
\bibitem [{\citenamefont {Liu}\ \emph {et~al.}(2020{\natexlab{a}})\citenamefont {Liu}, \citenamefont {Xie},\ and\ \citenamefont {Geng}}]{Liu:2020nil}%
  \BibitemOpen
  \bibfield  {author} {\bibinfo {author} {\bibfnamefont {M.-Z.}\ \bibnamefont {Liu}}, \bibinfo {author} {\bibfnamefont {J.-J.}\ \bibnamefont {Xie}}, \ and\ \bibinfo {author} {\bibfnamefont {L.-S.}\ \bibnamefont {Geng}},\ }\href {\doibase 10.1103/PhysRevD.102.091502} {\bibfield  {journal} {\bibinfo  {journal} {Phys. Rev. D}\ }\textbf {\bibinfo {volume} {102}},\ \bibinfo {pages} {091502} (\bibinfo {year} {2020}{\natexlab{a}})},\ \Eprint {http://arxiv.org/abs/2008.07389} {arXiv:2008.07389 [hep-ph]} \BibitemShut {NoStop}%
\bibitem [{\citenamefont {Hu}\ \emph {et~al.}(2021)\citenamefont {Hu}, \citenamefont {Lao}, \citenamefont {Ling},\ and\ \citenamefont {Wang}}]{Hu:2020mxp}%
  \BibitemOpen
  \bibfield  {author} {\bibinfo {author} {\bibfnamefont {M.-W.}\ \bibnamefont {Hu}}, \bibinfo {author} {\bibfnamefont {X.-Y.}\ \bibnamefont {Lao}}, \bibinfo {author} {\bibfnamefont {P.}~\bibnamefont {Ling}}, \ and\ \bibinfo {author} {\bibfnamefont {Q.}~\bibnamefont {Wang}},\ }\href {\doibase 10.1088/1674-1137/abcfaa} {\bibfield  {journal} {\bibinfo  {journal} {Chin. Phys. C}\ }\textbf {\bibinfo {volume} {45}},\ \bibinfo {pages} {021003} (\bibinfo {year} {2021})},\ \Eprint {http://arxiv.org/abs/2008.06894} {arXiv:2008.06894 [hep-ph]} \BibitemShut {NoStop}%
\bibitem [{\citenamefont {Agaev}\ \emph {et~al.}(2021)\citenamefont {Agaev}, \citenamefont {Azizi},\ and\ \citenamefont {Sundu}}]{Agaev:2020nrc}%
  \BibitemOpen
  \bibfield  {author} {\bibinfo {author} {\bibfnamefont {S.~S.}\ \bibnamefont {Agaev}}, \bibinfo {author} {\bibfnamefont {K.}~\bibnamefont {Azizi}}, \ and\ \bibinfo {author} {\bibfnamefont {H.}~\bibnamefont {Sundu}},\ }\href {\doibase 10.1088/1361-6471/ac0b31} {\bibfield  {journal} {\bibinfo  {journal} {J. Phys. G}\ }\textbf {\bibinfo {volume} {48}},\ \bibinfo {pages} {085012} (\bibinfo {year} {2021})},\ \Eprint {http://arxiv.org/abs/2008.13027} {arXiv:2008.13027 [hep-ph]} \BibitemShut {NoStop}%
\bibitem [{\citenamefont {Wang}\ and\ \citenamefont {Zhu}(2022)}]{Wang:2021lwy}%
  \BibitemOpen
  \bibfield  {author} {\bibinfo {author} {\bibfnamefont {B.}~\bibnamefont {Wang}}\ and\ \bibinfo {author} {\bibfnamefont {S.-L.}\ \bibnamefont {Zhu}},\ }\href {\doibase 10.1140/epjc/s10052-022-10396-9} {\bibfield  {journal} {\bibinfo  {journal} {Eur. Phys. J. C}\ }\textbf {\bibinfo {volume} {82}},\ \bibinfo {pages} {419} (\bibinfo {year} {2022})},\ \Eprint {http://arxiv.org/abs/2107.09275} {arXiv:2107.09275 [hep-ph]} \BibitemShut {NoStop}%
\bibitem [{\citenamefont {Ortega}\ \emph {et~al.}(2023)\citenamefont {Ortega}, \citenamefont {Entem}, \citenamefont {Fernandez},\ and\ \citenamefont {Segovia}}]{Ortega:2023azl}%
  \BibitemOpen
  \bibfield  {author} {\bibinfo {author} {\bibfnamefont {P.~G.}\ \bibnamefont {Ortega}}, \bibinfo {author} {\bibfnamefont {D.~R.}\ \bibnamefont {Entem}}, \bibinfo {author} {\bibfnamefont {F.}~\bibnamefont {Fernandez}}, \ and\ \bibinfo {author} {\bibfnamefont {J.}~\bibnamefont {Segovia}},\ }\href@noop {} {\  (\bibinfo {year} {2023})},\ \Eprint {http://arxiv.org/abs/2305.14430} {arXiv:2305.14430 [hep-ph]} \BibitemShut {NoStop}%
\bibitem [{\citenamefont {Wang}\ \emph {et~al.}(2024{\natexlab{a}})\citenamefont {Wang}, \citenamefont {Chen}, \citenamefont {Meng},\ and\ \citenamefont {Zhu}}]{Wang:2023hpp}%
  \BibitemOpen
  \bibfield  {author} {\bibinfo {author} {\bibfnamefont {B.}~\bibnamefont {Wang}}, \bibinfo {author} {\bibfnamefont {K.}~\bibnamefont {Chen}}, \bibinfo {author} {\bibfnamefont {L.}~\bibnamefont {Meng}}, \ and\ \bibinfo {author} {\bibfnamefont {S.-L.}\ \bibnamefont {Zhu}},\ }\href {\doibase 10.1103/PhysRevD.109.034027} {\bibfield  {journal} {\bibinfo  {journal} {Phys. Rev. D}\ }\textbf {\bibinfo {volume} {109}},\ \bibinfo {pages} {034027} (\bibinfo {year} {2024}{\natexlab{a}})},\ \Eprint {http://arxiv.org/abs/2309.02191} {arXiv:2309.02191 [hep-ph]} \BibitemShut {NoStop}%
\bibitem [{\citenamefont {Chen}\ \emph {et~al.}(2024)\citenamefont {Chen}, \citenamefont {Wu}, \citenamefont {Meng},\ and\ \citenamefont {Zhu}}]{Chen:2023syh}%
  \BibitemOpen
  \bibfield  {author} {\bibinfo {author} {\bibfnamefont {Y.-K.}\ \bibnamefont {Chen}}, \bibinfo {author} {\bibfnamefont {W.-L.}\ \bibnamefont {Wu}}, \bibinfo {author} {\bibfnamefont {L.}~\bibnamefont {Meng}}, \ and\ \bibinfo {author} {\bibfnamefont {S.-L.}\ \bibnamefont {Zhu}},\ }\href {\doibase 10.1103/PhysRevD.109.014010} {\bibfield  {journal} {\bibinfo  {journal} {Phys. Rev. D}\ }\textbf {\bibinfo {volume} {109}},\ \bibinfo {pages} {014010} (\bibinfo {year} {2024})},\ \Eprint {http://arxiv.org/abs/2310.14597} {arXiv:2310.14597 [hep-ph]} \BibitemShut {NoStop}%
\bibitem [{\citenamefont {Karliner}\ and\ \citenamefont {Rosner}(2020)}]{Karliner:2020vsi}%
  \BibitemOpen
  \bibfield  {author} {\bibinfo {author} {\bibfnamefont {M.}~\bibnamefont {Karliner}}\ and\ \bibinfo {author} {\bibfnamefont {J.~L.}\ \bibnamefont {Rosner}},\ }\href {\doibase 10.1103/PhysRevD.102.094016} {\bibfield  {journal} {\bibinfo  {journal} {Phys. Rev. D}\ }\textbf {\bibinfo {volume} {102}},\ \bibinfo {pages} {094016} (\bibinfo {year} {2020})},\ \Eprint {http://arxiv.org/abs/2008.05993} {arXiv:2008.05993 [hep-ph]} \BibitemShut {NoStop}%
\bibitem [{\citenamefont {He}\ \emph {et~al.}(2020)\citenamefont {He}, \citenamefont {Wang},\ and\ \citenamefont {Zhu}}]{He:2020jna}%
  \BibitemOpen
  \bibfield  {author} {\bibinfo {author} {\bibfnamefont {X.-G.}\ \bibnamefont {He}}, \bibinfo {author} {\bibfnamefont {W.}~\bibnamefont {Wang}}, \ and\ \bibinfo {author} {\bibfnamefont {R.}~\bibnamefont {Zhu}},\ }\href {\doibase 10.1140/epjc/s10052-020-08597-1} {\bibfield  {journal} {\bibinfo  {journal} {Eur. Phys. J. C}\ }\textbf {\bibinfo {volume} {80}},\ \bibinfo {pages} {1026} (\bibinfo {year} {2020})},\ \Eprint {http://arxiv.org/abs/2008.07145} {arXiv:2008.07145 [hep-ph]} \BibitemShut {NoStop}%
\bibitem [{\citenamefont {Wang}(2020)}]{Wang:2020xyc}%
  \BibitemOpen
  \bibfield  {author} {\bibinfo {author} {\bibfnamefont {Z.-G.}\ \bibnamefont {Wang}},\ }\href {\doibase 10.1142/S0217751X20501870} {\bibfield  {journal} {\bibinfo  {journal} {Int. J. Mod. Phys. A}\ }\textbf {\bibinfo {volume} {35}},\ \bibinfo {pages} {2050187} (\bibinfo {year} {2020})},\ \Eprint {http://arxiv.org/abs/2008.07833} {arXiv:2008.07833 [hep-ph]} \BibitemShut {NoStop}%
\bibitem [{\citenamefont {Zhang}(2021)}]{Zhang:2020oze}%
  \BibitemOpen
  \bibfield  {author} {\bibinfo {author} {\bibfnamefont {J.-R.}\ \bibnamefont {Zhang}},\ }\href {\doibase 10.1103/PhysRevD.103.054019} {\bibfield  {journal} {\bibinfo  {journal} {Phys. Rev. D}\ }\textbf {\bibinfo {volume} {103}},\ \bibinfo {pages} {054019} (\bibinfo {year} {2021})},\ \Eprint {http://arxiv.org/abs/2008.07295} {arXiv:2008.07295 [hep-ph]} \BibitemShut {NoStop}%
\bibitem [{\citenamefont {Wang}\ \emph {et~al.}(2021)\citenamefont {Wang}, \citenamefont {Meng}, \citenamefont {Xiao}, \citenamefont {Oka},\ and\ \citenamefont {Zhu}}]{Wang:2020prk}%
  \BibitemOpen
  \bibfield  {author} {\bibinfo {author} {\bibfnamefont {G.-J.}\ \bibnamefont {Wang}}, \bibinfo {author} {\bibfnamefont {L.}~\bibnamefont {Meng}}, \bibinfo {author} {\bibfnamefont {L.-Y.}\ \bibnamefont {Xiao}}, \bibinfo {author} {\bibfnamefont {M.}~\bibnamefont {Oka}}, \ and\ \bibinfo {author} {\bibfnamefont {S.-L.}\ \bibnamefont {Zhu}},\ }\href {\doibase 10.1140/epjc/s10052-021-08978-0} {\bibfield  {journal} {\bibinfo  {journal} {Eur. Phys. J. C}\ }\textbf {\bibinfo {volume} {81}},\ \bibinfo {pages} {188} (\bibinfo {year} {2021})},\ \Eprint {http://arxiv.org/abs/2010.09395} {arXiv:2010.09395 [hep-ph]} \BibitemShut {NoStop}%
\bibitem [{\citenamefont {L\"u}\ \emph {et~al.}(2020)\citenamefont {L\"u}, \citenamefont {Chen},\ and\ \citenamefont {Dong}}]{Lu:2020qmp}%
  \BibitemOpen
  \bibfield  {author} {\bibinfo {author} {\bibfnamefont {Q.-F.}\ \bibnamefont {L\"u}}, \bibinfo {author} {\bibfnamefont {D.-Y.}\ \bibnamefont {Chen}}, \ and\ \bibinfo {author} {\bibfnamefont {Y.-B.}\ \bibnamefont {Dong}},\ }\href {\doibase 10.1103/PhysRevD.102.074021} {\bibfield  {journal} {\bibinfo  {journal} {Phys. Rev. D}\ }\textbf {\bibinfo {volume} {102}},\ \bibinfo {pages} {074021} (\bibinfo {year} {2020})},\ \Eprint {http://arxiv.org/abs/2008.07340} {arXiv:2008.07340 [hep-ph]} \BibitemShut {NoStop}%
\bibitem [{\citenamefont {Tan}\ and\ \citenamefont {Ping}(2021)}]{Tan:2020cpu}%
  \BibitemOpen
  \bibfield  {author} {\bibinfo {author} {\bibfnamefont {Y.}~\bibnamefont {Tan}}\ and\ \bibinfo {author} {\bibfnamefont {J.}~\bibnamefont {Ping}},\ }\href {\doibase 10.1088/1674-1137/ac0ba4} {\bibfield  {journal} {\bibinfo  {journal} {Chin. Phys. C}\ }\textbf {\bibinfo {volume} {45}},\ \bibinfo {pages} {093104} (\bibinfo {year} {2021})},\ \Eprint {http://arxiv.org/abs/2010.04045} {arXiv:2010.04045 [hep-ph]} \BibitemShut {NoStop}%
\bibitem [{\citenamefont {Albuquerque}\ \emph {et~al.}(2021)\citenamefont {Albuquerque}, \citenamefont {Narison}, \citenamefont {Rabetiarivony},\ and\ \citenamefont {Randriamanatrika}}]{Albuquerque:2020ugi}%
  \BibitemOpen
  \bibfield  {author} {\bibinfo {author} {\bibfnamefont {R.~M.}\ \bibnamefont {Albuquerque}}, \bibinfo {author} {\bibfnamefont {S.}~\bibnamefont {Narison}}, \bibinfo {author} {\bibfnamefont {D.}~\bibnamefont {Rabetiarivony}}, \ and\ \bibinfo {author} {\bibfnamefont {G.}~\bibnamefont {Randriamanatrika}},\ }\href {\doibase 10.1016/j.nuclphysa.2020.122113} {\bibfield  {journal} {\bibinfo  {journal} {Nucl. Phys. A}\ }\textbf {\bibinfo {volume} {1007}},\ \bibinfo {pages} {122113} (\bibinfo {year} {2021})},\ \Eprint {http://arxiv.org/abs/2008.13463} {arXiv:2008.13463 [hep-ph]} \BibitemShut {NoStop}%
\bibitem [{\citenamefont {Agaev}\ \emph {et~al.}(2022)\citenamefont {Agaev}, \citenamefont {Azizi},\ and\ \citenamefont {Sundu}}]{Agaev:2022eeh}%
  \BibitemOpen
  \bibfield  {author} {\bibinfo {author} {\bibfnamefont {S.~S.}\ \bibnamefont {Agaev}}, \bibinfo {author} {\bibfnamefont {K.}~\bibnamefont {Azizi}}, \ and\ \bibinfo {author} {\bibfnamefont {H.}~\bibnamefont {Sundu}},\ }\href {\doibase 10.1103/PhysRevD.106.014019} {\bibfield  {journal} {\bibinfo  {journal} {Phys. Rev. D}\ }\textbf {\bibinfo {volume} {106}},\ \bibinfo {pages} {014019} (\bibinfo {year} {2022})},\ \Eprint {http://arxiv.org/abs/2204.08498} {arXiv:2204.08498 [hep-ph]} \BibitemShut {NoStop}%
\bibitem [{\citenamefont {Liu}\ \emph {et~al.}(2020{\natexlab{b}})\citenamefont {Liu}, \citenamefont {Yan}, \citenamefont {Ke}, \citenamefont {Li},\ and\ \citenamefont {Xie}}]{Liu:2020orv}%
  \BibitemOpen
  \bibfield  {author} {\bibinfo {author} {\bibfnamefont {X.-H.}\ \bibnamefont {Liu}}, \bibinfo {author} {\bibfnamefont {M.-J.}\ \bibnamefont {Yan}}, \bibinfo {author} {\bibfnamefont {H.-W.}\ \bibnamefont {Ke}}, \bibinfo {author} {\bibfnamefont {G.}~\bibnamefont {Li}}, \ and\ \bibinfo {author} {\bibfnamefont {J.-J.}\ \bibnamefont {Xie}},\ }\href {\doibase 10.1140/epjc/s10052-020-08762-6} {\bibfield  {journal} {\bibinfo  {journal} {Eur. Phys. J. C}\ }\textbf {\bibinfo {volume} {80}},\ \bibinfo {pages} {1178} (\bibinfo {year} {2020}{\natexlab{b}})},\ \Eprint {http://arxiv.org/abs/2008.07190} {arXiv:2008.07190 [hep-ph]} \BibitemShut {NoStop}%
\bibitem [{\citenamefont {Burns}\ and\ \citenamefont {Swanson}(2021)}]{Burns:2020epm}%
  \BibitemOpen
  \bibfield  {author} {\bibinfo {author} {\bibfnamefont {T.~J.}\ \bibnamefont {Burns}}\ and\ \bibinfo {author} {\bibfnamefont {E.~S.}\ \bibnamefont {Swanson}},\ }\href {\doibase 10.1016/j.physletb.2020.136057} {\bibfield  {journal} {\bibinfo  {journal} {Phys. Lett. B}\ }\textbf {\bibinfo {volume} {813}},\ \bibinfo {pages} {136057} (\bibinfo {year} {2021})},\ \Eprint {http://arxiv.org/abs/2008.12838} {arXiv:2008.12838 [hep-ph]} \BibitemShut {NoStop}%
\bibitem [{\citenamefont {Aaij}\ \emph {et~al.}(2024)\citenamefont {Aaij} \emph {et~al.}}]{LHCb:2024vfz}%
  \BibitemOpen
  \bibfield  {author} {\bibinfo {author} {\bibfnamefont {R.}~\bibnamefont {Aaij}} \emph {et~al.} (\bibinfo {collaboration} {LHCb}),\ }\href@noop {} {\  (\bibinfo {year} {2024})},\ \Eprint {http://arxiv.org/abs/2406.03156} {arXiv:2406.03156 [hep-ex]} \BibitemShut {NoStop}%
\bibitem [{\citenamefont {Molina}\ \emph {et~al.}(2010)\citenamefont {Molina}, \citenamefont {Branz},\ and\ \citenamefont {Oset}}]{Molina:2010tx}%
  \BibitemOpen
  \bibfield  {author} {\bibinfo {author} {\bibfnamefont {R.}~\bibnamefont {Molina}}, \bibinfo {author} {\bibfnamefont {T.}~\bibnamefont {Branz}}, \ and\ \bibinfo {author} {\bibfnamefont {E.}~\bibnamefont {Oset}},\ }\href {\doibase 10.1103/PhysRevD.82.014010} {\bibfield  {journal} {\bibinfo  {journal} {Phys. Rev. D}\ }\textbf {\bibinfo {volume} {82}},\ \bibinfo {pages} {014010} (\bibinfo {year} {2010})},\ \Eprint {http://arxiv.org/abs/1005.0335} {arXiv:1005.0335 [hep-ph]} \BibitemShut {NoStop}%
\bibitem [{\citenamefont {Lin}\ \emph {et~al.}(2024)\citenamefont {Lin}, \citenamefont {Wang}, \citenamefont {Cheng}, \citenamefont {Meng},\ and\ \citenamefont {Zhu}}]{Lin:2024qcq}%
  \BibitemOpen
  \bibfield  {author} {\bibinfo {author} {\bibfnamefont {Z.-Y.}\ \bibnamefont {Lin}}, \bibinfo {author} {\bibfnamefont {J.-Z.}\ \bibnamefont {Wang}}, \bibinfo {author} {\bibfnamefont {J.-B.}\ \bibnamefont {Cheng}}, \bibinfo {author} {\bibfnamefont {L.}~\bibnamefont {Meng}}, \ and\ \bibinfo {author} {\bibfnamefont {S.-L.}\ \bibnamefont {Zhu}},\ }\href@noop {} {\  (\bibinfo {year} {2024})},\ \Eprint {http://arxiv.org/abs/2403.01727} {arXiv:2403.01727 [hep-ph]} \BibitemShut {NoStop}%
\bibitem [{\citenamefont {Ablikim}\ \emph {et~al.}(2024)\citenamefont {Ablikim} \emph {et~al.}}]{BESIII:2024ths}%
  \BibitemOpen
  \bibfield  {author} {\bibinfo {author} {\bibfnamefont {M.}~\bibnamefont {Ablikim}} \emph {et~al.} (\bibinfo {collaboration} {BESIII}),\ }\href@noop {} {\  (\bibinfo {year} {2024})},\ \Eprint {http://arxiv.org/abs/2402.03829} {arXiv:2402.03829 [hep-ex]} \BibitemShut {NoStop}%
\bibitem [{\citenamefont {Silvestre-Brac}(1996)}]{Silvestre-Brac:1996myf}%
  \BibitemOpen
  \bibfield  {author} {\bibinfo {author} {\bibfnamefont {B.}~\bibnamefont {Silvestre-Brac}},\ }\href {\doibase 10.1007/s006010050028} {\bibfield  {journal} {\bibinfo  {journal} {Few Body Syst.}\ }\textbf {\bibinfo {volume} {20}},\ \bibinfo {pages} {1} (\bibinfo {year} {1996})}\BibitemShut {NoStop}%
\bibitem [{\citenamefont {Vijande}\ \emph {et~al.}(2005)\citenamefont {Vijande}, \citenamefont {Fernandez},\ and\ \citenamefont {Valcarce}}]{Vijande:2004he}%
  \BibitemOpen
  \bibfield  {author} {\bibinfo {author} {\bibfnamefont {J.}~\bibnamefont {Vijande}}, \bibinfo {author} {\bibfnamefont {F.}~\bibnamefont {Fernandez}}, \ and\ \bibinfo {author} {\bibfnamefont {A.}~\bibnamefont {Valcarce}},\ }\href {\doibase 10.1088/0954-3899/31/5/017} {\bibfield  {journal} {\bibinfo  {journal} {J. Phys. G}\ }\textbf {\bibinfo {volume} {31}},\ \bibinfo {pages} {481} (\bibinfo {year} {2005})},\ \Eprint {http://arxiv.org/abs/hep-ph/0411299} {arXiv:hep-ph/0411299} \BibitemShut {NoStop}%
\bibitem [{\citenamefont {Wise}(1992)}]{Wise:1992hn}%
  \BibitemOpen
  \bibfield  {author} {\bibinfo {author} {\bibfnamefont {M.~B.}\ \bibnamefont {Wise}},\ }\href {\doibase 10.1103/PhysRevD.45.R2188} {\bibfield  {journal} {\bibinfo  {journal} {Phys. Rev. D}\ }\textbf {\bibinfo {volume} {45}},\ \bibinfo {pages} {R2188} (\bibinfo {year} {1992})}\BibitemShut {NoStop}%
\bibitem [{\citenamefont {Yan}\ \emph {et~al.}(1992)\citenamefont {Yan}, \citenamefont {Cheng}, \citenamefont {Cheung}, \citenamefont {Lin}, \citenamefont {Lin},\ and\ \citenamefont {Yu}}]{Yan:1992gz}%
  \BibitemOpen
  \bibfield  {author} {\bibinfo {author} {\bibfnamefont {T.-M.}\ \bibnamefont {Yan}}, \bibinfo {author} {\bibfnamefont {H.-Y.}\ \bibnamefont {Cheng}}, \bibinfo {author} {\bibfnamefont {C.-Y.}\ \bibnamefont {Cheung}}, \bibinfo {author} {\bibfnamefont {G.-L.}\ \bibnamefont {Lin}}, \bibinfo {author} {\bibfnamefont {Y.~C.}\ \bibnamefont {Lin}}, \ and\ \bibinfo {author} {\bibfnamefont {H.-L.}\ \bibnamefont {Yu}},\ }\href {\doibase 10.1103/PhysRevD.46.1148} {\bibfield  {journal} {\bibinfo  {journal} {Phys. Rev. D}\ }\textbf {\bibinfo {volume} {46}},\ \bibinfo {pages} {1148} (\bibinfo {year} {1992})},\ \bibinfo {note} {[Erratum: Phys.Rev.D 55, 5851 (1997)]}\BibitemShut {NoStop}%
\bibitem [{\citenamefont {Grinstein}\ \emph {et~al.}(1992)\citenamefont {Grinstein}, \citenamefont {Jenkins}, \citenamefont {Manohar}, \citenamefont {Savage},\ and\ \citenamefont {Wise}}]{Grinstein:1992qt}%
  \BibitemOpen
  \bibfield  {author} {\bibinfo {author} {\bibfnamefont {B.}~\bibnamefont {Grinstein}}, \bibinfo {author} {\bibfnamefont {E.~E.}\ \bibnamefont {Jenkins}}, \bibinfo {author} {\bibfnamefont {A.~V.}\ \bibnamefont {Manohar}}, \bibinfo {author} {\bibfnamefont {M.~J.}\ \bibnamefont {Savage}}, \ and\ \bibinfo {author} {\bibfnamefont {M.~B.}\ \bibnamefont {Wise}},\ }\href {\doibase 10.1016/0550-3213(92)90248-A} {\bibfield  {journal} {\bibinfo  {journal} {Nucl. Phys. B}\ }\textbf {\bibinfo {volume} {380}},\ \bibinfo {pages} {369} (\bibinfo {year} {1992})},\ \Eprint {http://arxiv.org/abs/hep-ph/9204207} {arXiv:hep-ph/9204207} \BibitemShut {NoStop}%
\bibitem [{\citenamefont {Casalbuoni}\ \emph {et~al.}(1997)\citenamefont {Casalbuoni}, \citenamefont {Deandrea}, \citenamefont {Di~Bartolomeo}, \citenamefont {Gatto}, \citenamefont {Feruglio},\ and\ \citenamefont {Nardulli}}]{Casalbuoni:1996pg}%
  \BibitemOpen
  \bibfield  {author} {\bibinfo {author} {\bibfnamefont {R.}~\bibnamefont {Casalbuoni}}, \bibinfo {author} {\bibfnamefont {A.}~\bibnamefont {Deandrea}}, \bibinfo {author} {\bibfnamefont {N.}~\bibnamefont {Di~Bartolomeo}}, \bibinfo {author} {\bibfnamefont {R.}~\bibnamefont {Gatto}}, \bibinfo {author} {\bibfnamefont {F.}~\bibnamefont {Feruglio}}, \ and\ \bibinfo {author} {\bibfnamefont {G.}~\bibnamefont {Nardulli}},\ }\href {\doibase 10.1016/S0370-1573(96)00027-0} {\bibfield  {journal} {\bibinfo  {journal} {Phys. Rept.}\ }\textbf {\bibinfo {volume} {281}},\ \bibinfo {pages} {145} (\bibinfo {year} {1997})},\ \Eprint {http://arxiv.org/abs/hep-ph/9605342} {arXiv:hep-ph/9605342} \BibitemShut {NoStop}%
\bibitem [{\citenamefont {Li}\ and\ \citenamefont {Zhu}(2012)}]{Li:2012cs}%
  \BibitemOpen
  \bibfield  {author} {\bibinfo {author} {\bibfnamefont {N.}~\bibnamefont {Li}}\ and\ \bibinfo {author} {\bibfnamefont {S.-L.}\ \bibnamefont {Zhu}},\ }\href {\doibase 10.1103/PhysRevD.86.074022} {\bibfield  {journal} {\bibinfo  {journal} {Phys. Rev. D}\ }\textbf {\bibinfo {volume} {86}},\ \bibinfo {pages} {074022} (\bibinfo {year} {2012})},\ \Eprint {http://arxiv.org/abs/1207.3954} {arXiv:1207.3954 [hep-ph]} \BibitemShut {NoStop}%
\bibitem [{\citenamefont {Li}\ \emph {et~al.}(2013)\citenamefont {Li}, \citenamefont {Sun}, \citenamefont {Liu},\ and\ \citenamefont {Zhu}}]{Li:2012ss}%
  \BibitemOpen
  \bibfield  {author} {\bibinfo {author} {\bibfnamefont {N.}~\bibnamefont {Li}}, \bibinfo {author} {\bibfnamefont {Z.-F.}\ \bibnamefont {Sun}}, \bibinfo {author} {\bibfnamefont {X.}~\bibnamefont {Liu}}, \ and\ \bibinfo {author} {\bibfnamefont {S.-L.}\ \bibnamefont {Zhu}},\ }\href {\doibase 10.1103/PhysRevD.88.114008} {\bibfield  {journal} {\bibinfo  {journal} {Phys. Rev. D}\ }\textbf {\bibinfo {volume} {88}},\ \bibinfo {pages} {114008} (\bibinfo {year} {2013})},\ \Eprint {http://arxiv.org/abs/1211.5007} {arXiv:1211.5007 [hep-ph]} \BibitemShut {NoStop}%
\bibitem [{\citenamefont {Isola}\ \emph {et~al.}(2003)\citenamefont {Isola}, \citenamefont {Ladisa}, \citenamefont {Nardulli},\ and\ \citenamefont {Santorelli}}]{Isola:2003fh}%
  \BibitemOpen
  \bibfield  {author} {\bibinfo {author} {\bibfnamefont {C.}~\bibnamefont {Isola}}, \bibinfo {author} {\bibfnamefont {M.}~\bibnamefont {Ladisa}}, \bibinfo {author} {\bibfnamefont {G.}~\bibnamefont {Nardulli}}, \ and\ \bibinfo {author} {\bibfnamefont {P.}~\bibnamefont {Santorelli}},\ }\href {\doibase 10.1103/PhysRevD.68.114001} {\bibfield  {journal} {\bibinfo  {journal} {Phys. Rev. D}\ }\textbf {\bibinfo {volume} {68}},\ \bibinfo {pages} {114001} (\bibinfo {year} {2003})},\ \Eprint {http://arxiv.org/abs/hep-ph/0307367} {arXiv:hep-ph/0307367} \BibitemShut {NoStop}%
\bibitem [{\citenamefont {Bando}\ \emph {et~al.}(1988)\citenamefont {Bando}, \citenamefont {Kugo},\ and\ \citenamefont {Yamawaki}}]{Bando:1987br}%
  \BibitemOpen
  \bibfield  {author} {\bibinfo {author} {\bibfnamefont {M.}~\bibnamefont {Bando}}, \bibinfo {author} {\bibfnamefont {T.}~\bibnamefont {Kugo}}, \ and\ \bibinfo {author} {\bibfnamefont {K.}~\bibnamefont {Yamawaki}},\ }\href {\doibase 10.1016/0370-1573(88)90019-1} {\bibfield  {journal} {\bibinfo  {journal} {Phys. Rept.}\ }\textbf {\bibinfo {volume} {164}},\ \bibinfo {pages} {217} (\bibinfo {year} {1988})}\BibitemShut {NoStop}%
\bibitem [{\citenamefont {Liu}\ \emph {et~al.}(2008)\citenamefont {Liu}, \citenamefont {Liu}, \citenamefont {Deng},\ and\ \citenamefont {Zhu}}]{Liu:2008xz}%
  \BibitemOpen
  \bibfield  {author} {\bibinfo {author} {\bibfnamefont {X.}~\bibnamefont {Liu}}, \bibinfo {author} {\bibfnamefont {Y.-R.}\ \bibnamefont {Liu}}, \bibinfo {author} {\bibfnamefont {W.-Z.}\ \bibnamefont {Deng}}, \ and\ \bibinfo {author} {\bibfnamefont {S.-L.}\ \bibnamefont {Zhu}},\ }\href {\doibase 10.1103/PhysRevD.77.094015} {\bibfield  {journal} {\bibinfo  {journal} {Phys. Rev. D}\ }\textbf {\bibinfo {volume} {77}},\ \bibinfo {pages} {094015} (\bibinfo {year} {2008})},\ \Eprint {http://arxiv.org/abs/0803.1295} {arXiv:0803.1295 [hep-ph]} \BibitemShut {NoStop}%
\bibitem [{\citenamefont {Aguilar}\ and\ \citenamefont {Combes}(1971)}]{Aguilar:1971ve}%
  \BibitemOpen
  \bibfield  {author} {\bibinfo {author} {\bibfnamefont {J.}~\bibnamefont {Aguilar}}\ and\ \bibinfo {author} {\bibfnamefont {J.~M.}\ \bibnamefont {Combes}},\ }\href {\doibase 10.1007/BF01877510} {\bibfield  {journal} {\bibinfo  {journal} {Commun. Math. Phys.}\ }\textbf {\bibinfo {volume} {22}},\ \bibinfo {pages} {269} (\bibinfo {year} {1971})}\BibitemShut {NoStop}%
\bibitem [{\citenamefont {Balslev}\ and\ \citenamefont {Combes}(1971)}]{Balslev:1971vb}%
  \BibitemOpen
  \bibfield  {author} {\bibinfo {author} {\bibfnamefont {E.}~\bibnamefont {Balslev}}\ and\ \bibinfo {author} {\bibfnamefont {J.~M.}\ \bibnamefont {Combes}},\ }\href {\doibase 10.1007/BF01877511} {\bibfield  {journal} {\bibinfo  {journal} {Commun. Math. Phys.}\ }\textbf {\bibinfo {volume} {22}},\ \bibinfo {pages} {280} (\bibinfo {year} {1971})}\BibitemShut {NoStop}%
\bibitem [{\citenamefont {Lin}\ \emph {et~al.}(2022)\citenamefont {Lin}, \citenamefont {Cheng},\ and\ \citenamefont {Zhu}}]{Lin:2022wmj}%
  \BibitemOpen
  \bibfield  {author} {\bibinfo {author} {\bibfnamefont {Z.-Y.}\ \bibnamefont {Lin}}, \bibinfo {author} {\bibfnamefont {J.-B.}\ \bibnamefont {Cheng}}, \ and\ \bibinfo {author} {\bibfnamefont {S.-L.}\ \bibnamefont {Zhu}},\ }\href@noop {} {\  (\bibinfo {year} {2022})},\ \Eprint {http://arxiv.org/abs/2205.14628} {arXiv:2205.14628 [hep-ph]} \BibitemShut {NoStop}%
\bibitem [{\citenamefont {Lin}\ \emph {et~al.}(2023)\citenamefont {Lin}, \citenamefont {Cheng}, \citenamefont {Huang},\ and\ \citenamefont {Zhu}}]{Lin:2023ihj}%
  \BibitemOpen
  \bibfield  {author} {\bibinfo {author} {\bibfnamefont {Z.-Y.}\ \bibnamefont {Lin}}, \bibinfo {author} {\bibfnamefont {J.-B.}\ \bibnamefont {Cheng}}, \bibinfo {author} {\bibfnamefont {B.-L.}\ \bibnamefont {Huang}}, \ and\ \bibinfo {author} {\bibfnamefont {S.-L.}\ \bibnamefont {Zhu}},\ }\href {\doibase 10.1103/PhysRevD.108.114014} {\bibfield  {journal} {\bibinfo  {journal} {Phys. Rev. D}\ }\textbf {\bibinfo {volume} {108}},\ \bibinfo {pages} {114014} (\bibinfo {year} {2023})},\ \Eprint {http://arxiv.org/abs/2305.19073} {arXiv:2305.19073 [hep-ph]} \BibitemShut {NoStop}%
\bibitem [{\citenamefont {Wang}\ \emph {et~al.}(2024{\natexlab{b}})\citenamefont {Wang}, \citenamefont {Lin}, \citenamefont {Chen}, \citenamefont {Meng},\ and\ \citenamefont {Zhu}}]{Wang:2024ytk}%
  \BibitemOpen
  \bibfield  {author} {\bibinfo {author} {\bibfnamefont {J.-Z.}\ \bibnamefont {Wang}}, \bibinfo {author} {\bibfnamefont {Z.-Y.}\ \bibnamefont {Lin}}, \bibinfo {author} {\bibfnamefont {Y.-K.}\ \bibnamefont {Chen}}, \bibinfo {author} {\bibfnamefont {L.}~\bibnamefont {Meng}}, \ and\ \bibinfo {author} {\bibfnamefont {S.-L.}\ \bibnamefont {Zhu}},\ }\href@noop {} {\  (\bibinfo {year} {2024}{\natexlab{b}})},\ \Eprint {http://arxiv.org/abs/2404.16575} {arXiv:2404.16575 [hep-ph]} \BibitemShut {NoStop}%
\bibitem [{\citenamefont {Shi}\ \emph {et~al.}(2024)\citenamefont {Shi}, \citenamefont {Gil-Dom\'\i{}nguez}, \citenamefont {Molina},\ and\ \citenamefont {Du}}]{Shi:2024squ}%
  \BibitemOpen
  \bibfield  {author} {\bibinfo {author} {\bibfnamefont {P.-P.}\ \bibnamefont {Shi}}, \bibinfo {author} {\bibfnamefont {F.}~\bibnamefont {Gil-Dom\'\i{}nguez}}, \bibinfo {author} {\bibfnamefont {R.}~\bibnamefont {Molina}}, \ and\ \bibinfo {author} {\bibfnamefont {M.-L.}\ \bibnamefont {Du}},\ }\href@noop {} {\  (\bibinfo {year} {2024})},\ \Eprint {http://arxiv.org/abs/2408.01197} {arXiv:2408.01197 [hep-ph]} \BibitemShut {NoStop}%
\bibitem [{\citenamefont {Aaij}\ \emph {et~al.}(2023{\natexlab{a}})\citenamefont {Aaij} \emph {et~al.}}]{LHCb:2022sfr}%
  \BibitemOpen
  \bibfield  {author} {\bibinfo {author} {\bibfnamefont {R.}~\bibnamefont {Aaij}} \emph {et~al.} (\bibinfo {collaboration} {LHCb}),\ }\href {\doibase 10.1103/PhysRevLett.131.041902} {\bibfield  {journal} {\bibinfo  {journal} {Phys. Rev. Lett.}\ }\textbf {\bibinfo {volume} {131}},\ \bibinfo {pages} {041902} (\bibinfo {year} {2023}{\natexlab{a}})},\ \Eprint {http://arxiv.org/abs/2212.02716} {arXiv:2212.02716 [hep-ex]} \BibitemShut {NoStop}%
\bibitem [{\citenamefont {Aaij}\ \emph {et~al.}(2023{\natexlab{b}})\citenamefont {Aaij} \emph {et~al.}}]{LHCb:2022lzp}%
  \BibitemOpen
  \bibfield  {author} {\bibinfo {author} {\bibfnamefont {R.}~\bibnamefont {Aaij}} \emph {et~al.} (\bibinfo {collaboration} {LHCb}),\ }\href {\doibase 10.1103/PhysRevD.108.012017} {\bibfield  {journal} {\bibinfo  {journal} {Phys. Rev. D}\ }\textbf {\bibinfo {volume} {108}},\ \bibinfo {pages} {012017} (\bibinfo {year} {2023}{\natexlab{b}})},\ \Eprint {http://arxiv.org/abs/2212.02717} {arXiv:2212.02717 [hep-ex]} \BibitemShut {NoStop}%
\end{thebibliography}%
\end{document}